\documentclass[oldversion]{aa} 
\usepackage{graphicx}
\usepackage{longtable}
\usepackage{txfonts}
\usepackage{rotating}
\usepackage{lscape}
\newcommand{\gsim}{\;\lower.6ex\hbox{$\sim$}\kern-7.75pt\raise.65ex\hbox{$>$}\;}
\newcommand{\lsim}{\;\lower.6ex\hbox{$\sim$}\kern-7.75pt\raise.65ex\hbox{$<$}\;}

\begin{document}
\title{Multiple stellar populations in the globular cluster 
NGC~1851\thanks{Based on observations collected at 
ESO telescopes under programme 083.D-0208}\fnmsep\thanks{
   Tables 2, 3, 4, 5, 6, 7 and 8 are only available in electronic form at the CDS via anonymous
   ftp to {\tt cdsarc.u-strasbg.fr} (130.79.128.5) or via
   {\tt http://cdsweb.u-strasbg.fr/cgi-bin/qcat?J/A+A/???/???}}
 }

\author{
E. Carretta\inst{1},
S. Lucatello\inst{2},
R.G. Gratton\inst{2},
A. Bragaglia\inst{1},
\and
V. D'Orazi\inst{2}.
}

\authorrunning{E. Carretta et al.}
\titlerunning{Multiple stellar populations in NGC~1851}

\offprints{E. Carretta, eugenio.carretta@oabo.inaf.it}

\institute{
INAF-Osservatorio Astronomico di Bologna, Via Ranzani 1, I-40127
 Bologna, Italy
\and
INAF-Osservatorio Astronomico di Padova, Vicolo dell'Osservatorio 5, I-35122
 Padova, Italy
  }

\date{}

\abstract{Detailed chemical tagging of individual stellar populations in the
Galactic globular cluster (GC) NGC~1851 is presented. Abundances are derived
from FLAMES spectra for the largest sample of giants (124) and the largest
number of elements ever analysed in this peculiar GC. The chemistry is
characterised using homogeneous abundances of proton-capture (O, Na, Mg, Al,
Si), $\alpha-$capture (Ca, Ti), Fe-peak (Sc, V, Mn, Co, Ni, Cu), and
neutron-capture elements (Y, Zr, Ba, La, Ce, Nd, Eu, Dy). We confirm the
presence of an [Fe/H] spread larger than observational errors in this cluster,
but too small to clearly separate different sub-populations. We instead
propose a classification scheme using a combination of Fe and Ba (which is much more abundant in the
more metal-rich group) through a cluster analysis. With this approach, we
separated stars into two components (the metal-rich (MR) and metal-poor (MP) 
populations), each of them showing a Na-O anticorrelation, signature of genuine 
GC, although with different ratios of primordial (FG) to polluted (SG) stars.
Moreover,  clear (anti)correlations of Mg and Si with Na and O are found for
each component.  The level of [$\alpha$/H] tracks iron and is higher in the MR
population, suggesting an additional contribution by core-collapse
supernovae. When considering all $s-$process elements, the MR
population shows a larger enrichment compared to the MP one.
This is probably due to the  contribution by intermediate-low
mass stars, because we find that the level of  heavy $s-$process elements is
larger than the level of light $s-$process nuclei in the MR stars; however, a
large contribution by low mass stars is unlikely,  because it would likely
cancel the O-Na anticorrelation.  Finally, we confirm the presence of
correlations between the amount  of proton-capture elements and the level of
$s-$process elements previously found by other investigations, at least for the
MR population. This finding apparently  requires a quite long delay for the
second generation of the MR component. Scenarios for the formation of NGC~1851
appear complex, and are not yet well understood. A merger of two distinct GCs in
a parent dwarf galaxy, each cluster with a different  Ba level and an age
difference of $\sim 1$~Gyr, might explain (i) the  double subgiant branch, (ii)
a possible difference in C content between the two original GCs, and (iii) the
Str\"omgren photometry of this peculiar cluster. However, the correlation
existing between $p-$capture and $n-$capture elements within the MR population
requires the further assumption of a long delay for its second generation. More
observations are required to fully understand the formation of this GC.}

\maketitle

\section{Introduction}

Spectroscopy (starting several decades ago) and more recently photometry (in
particular in the near ultraviolet window) have shown that globular clusters (GCs)
are  much more complex than previously imagined. Since the publishing of the
pioneering work by Cohen (1978) and Peterson (1980), more than 30 years of
abundance analysis of GC stars have shown that they host multiple populations,
with slightly different ages but showing even very different chemical
composition (see Kraft 1994 and Gratton, Sneden \& Carretta 2004 for reviews and
references). {\it The most outstanding signature is provided by the observed
Na-O anticorrelation, detected so far in every GC investigated with high
resolution, high quality spectra}\footnote{In GCs in which  the
anti-correlation was not found, Pal12 and Ter 7
(Sbordone et al. 2007, Cohen 2004, respectively) only a few
stars were studied, hence small number statistics prevents the drawing of any
conclusion.}.  This feature is so widespread to be interpreted as a birthmark of
genuine, {\it bona fide} GCs (Carretta et al. 2010a), likely a
result of their very own formation mechanism(s) (Carretta 2006). Although the
exact chain of events leading to the GCs we actually observe is still in a
``tentative scenario" stage, some firm points are currently well assessed.

The enhancement in Na (and N, Al) and the depletion in O (and C, Mg) are
observed also among unevolved stars (e.g. Gratton et al. 2001), bearing evidence
of a previous generation of more massive stars ending their short lifetimes with
the injection of nuclearly processed matter in the intra-cluster medium. From
the mix of pristine material with these enriched ejecta (see e.g. Prantzos and
Charbonnel 2006) a second generation of stars is born. The typical shape of the
Na-O anticorrelation is then probably due to some dilution with pre-existing gas 
of the ashes from proton-capture reactions in H-burning at quite high temperatures
(Denisenkov \& Denisenkova 1989, Langer et al. 1993). The degree of the
processing is modulated by some combination of global cluster parameters, the 
critical ones being total mass and metallicity (Carretta et al. 2010a). This is
basically the scenario resulting from the operation of early polluters in the
first phases of cluster life, although the exact nature of these polluting stars
is still debated (massive, rapidly rotating stars, Decressin et al. 2007, $vs$
intermediate mass asymptotic branch AGB stars, Ventura et al. 2001, are currently 
the main candidates).

However, our ongoing extensive FLAMES survey of Na-O anticorrelation in more
than 20 GCs (see Carretta et al. 2006, 2009a,b,c, 2010b,c; Gratton 2010a for
strategy  and the updated status of the survey) also revealed that superimposed
to this common pattern, in some cases there is one (or more) further degree of
complexity. Even neglecting the cases of likely remnants of dispersed/dispersing
dwarf galaxies (like M~54 and possibly $\omega$~Cen, Carretta et al. 2010b),
the globular cluster NGC~1851 stands out.

NGC~1851 has been known for a long time to show a bimodal colour distribution of
stars in the core He-burning phase (e.g. Walker 1992): its horizontal branch
(HB) is discontinuous, with a well populated red HB (RHB), a few RR Lyrae stars
and an extended blue HB (BHB; Fig.~\ref{f:cmdsel1851}). Despite being somewhat
less massive than clusters showing distinct multiple main sequences (like
NGC~2808 and $\omega$~Cen, D'Antona et al. 2005 and Anderson 1998,
respectively), NGC~1851 was found to contain two subgiant branches (SGBs; Milone
et al. 2008) with a magnitude difference corresponding to about 1~Gyr, if
interpreted as an age difference. The same split may however be explained by a
spread of the total CNO abundances (Cassisi et al. 2008); such a spread was
claimed by Yong et al. (2009) to exist in NGC~1851 from the analysis of four
red  giant branch (RGB) stars, while a constant sum of C+N+O abundances was recently
derived for a sample of 15 RGB stars by Villanova et al. (2010, hereinafter V10).

The same level of luminosity shared by the RHB and the BHB seems to exclude a
large difference of He among stellar populations in NGC~1851 (Cassisi et al. 2008).
On the other hand, the lack of splitting in the MS implies a similar [Fe/H]\footnote{We adopt the
usual spectroscopic notation, $i.e.$  for any given species X, [X]= 
$\log{\epsilon(X)_{\rm star}} - \log{\epsilon(X)_\odot}$ and  
$\log{\epsilon(X)}=\log{(N_{\rm X}/N_{\rm H})}+12.0$\ for absolute number 
density abundances.} for the two sequences. However, this does not exclude 
the presence of a small metallicity spread. Both the population ratios and the
central concentration led some studies to connect the RHB to the younger SGB and
the extended BHB to the older SGB, in the framework of an age difference.
However, the issue of a different radial distribution was still debated until
recently (Zoccali et al. 2009; Milone et al. 2009). Our observations of a
small metallicity spread and of a different distribution of the most metal-poor 
and metal-rich component on the RGB (Carretta et al. 2010c) finally provided a 
conclusive evidence that two stellar populations, differently concentrated within 
the cluster, do co-exist in NGC~1851.

The finding of a small but we think real metallicity dispersion in NGC~1851 was 
immediately compared with the split of the RGB uncovered by Lee et
al. (2009a) using narrow band Ca~II photometry and by Han et al. (2009) with
$UVI$\ Johnson broadband photometry. In turn this might be interpreted in the
scenario devised by Lee et al. (2009b) who claimed an enrichment by supernovae
in GCs with multiple populations. While Carretta et al. (2010d) confuted the
general application of this contribution to {\it all} GCs, the reality of different
levels of Ca and other $\alpha-$elements from core-collapse SNe could not be
excluded for this particular cluster.

To make things still more complicate, Yong and Grundahl (2008) found variations
of Zr and La correlated with those of Al, suggesting some contribution from
thermally pulsing AGB stars. This introduces another problem since light
$s-$process elements and Al are thought to be produced by stars of different
mass ranges. When different mass ranges are involved, some time delay and/or
prolonged star formation must also be taken into account, increasing the level
of complexity in any possible modelling of the earliest phase of cluster
formation and evolution.

To deal with all the old and new pieces of the "puzzle NGC~1851", we tentatively
proposed in Carretta et al. (2010c) a scenario that may account for all the
observations there considered. According to this scenario, the current cluster
is the result of a merger between two distinct clusters born in a much larger
system, like a dwarf galaxy now no longer visible. A subset of elements was used
to show that this hypothesis, already put forward by some authors (van den Bergh
1996, Catelan 1997), is able to accommodate most of the observations. 

Here we present our entire large dataset  consisting in abundances for many 
species derived in more than 120 red giants in NGC~1851. This dataset will be 
discussed within this framework. The paper is organised as follow. Our data and 
analysis techniques are presented in Sections 2 and 3, with particular emphasis 
on neutron-capture elements, whose abundances play an important role in this GC
and were not considered in detail in Carretta et al. (2010c). 
The issues of the metallicity spread and of the operative definition of stellar
components in NGC~1851 are discussed in Section 4. The complex nucleosynthesis
observed in this GC is examined in detail in Section 5, while an extensive
comparison of results from spectroscopy and photometry is presented in Section 6.
Finally, these data are discussed in Section 7, where some tentative conclusion
will be drawn. The Appendix gives detail about the error analysis.

\section{Sample selection and observations}

\begin{figure}
\centering
\includegraphics[scale=0.44]{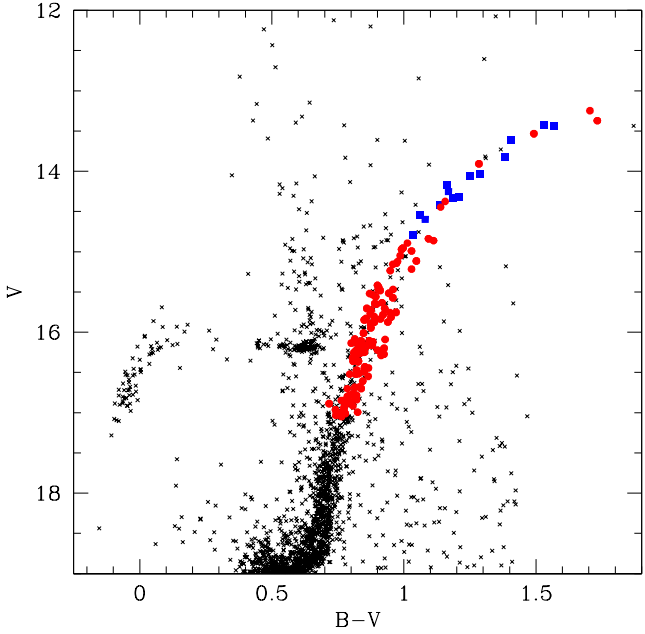}
\caption{The $V,B-V$ CMD of NGC~1851 with stars selected for the present study
indicated by large red filled circles (GIRAFFE targets) and large blue squares
(UVES targets).}
\label{f:cmdsel1851}
\end{figure}

The sample of RGB stars to be observed in NGC~1851 was selected with the same
criteria (small distance from RGB ridge line, absence of close companions)
adopted in previous works of our survey. FLAMES observations (the same used
also in Carretta et al. 2010c) consisted in three pointings with the HR11 high
resolution grating (R=24200) covering the Na~{\sc i} 5682-88~\AA\ doublet
and four pointings  with the grating HR13 (R=22500) including the [O~{\sc i}] forbidden lines at 6300-63~\AA . The median S/N ratio is about 75. The
observation log is listed in Tab.~\ref{t:logobs} and the selected stars are
shown on a $V,B-V$ colour-magnitude diagram (CMD) in Fig.~\ref{f:cmdsel1851}.

$V$\ and $B$\ band photometry (kindly provided by Y. Momany) from ESO Imaging
Survey frames was complemented with near infrared photometry, to obtain 
atmospheric parameters as described below. $K$ band magnitudes were 
obtained from  the Point Source Catalogue of 2MASS (Skrutskie et al. 2006) and 
properly transformed to the TCS photometric system, to be used to derive input 
atmospheric parameters, following the Alonso et al. (1999) calibration (see below).

Initially, 124 member stars (according to the radial velocity RV, see below)
were targeted with GIRAFFE fibres and 14 stars with the fibre-fed UVES 
spectrograph (Red Arm, with spectral range from 4800 to 6800~\AA and R=45000); 10 stars 
were observed with both instruments.

\begin{table}
\centering
\caption{Log of FLAMES observations for NGC 1851}
\begin{tabular}{cccccc}
\hline\hline
   Date         &     UT       & exp. & grating & seeing & airmass\\
                &              & (sec)&         &        &        \\ 
\hline
 Apr. 09, 2010  &23:53:06.585  &  2880   & HR11    & 0.82   & 1.355  \\
 Apr. 11, 2010  &00:21:43.120  &  2880   & HR11    & 0.67   & 1.514  \\
 Aug. 05, 2010  &09:13:30.637  &  2880   & HR11    & 1.02   & 1.533  \\
 Aug. 26, 2010  &07:57:12.771  &  2880   & HR13    & 1.21   & 1.497  \\
 Aug. 26, 2010  &08:47:34.969  &  2880   & HR13    & 1.11   & 1.278  \\
 Sep. 01, 2010  &08:00:05.991  &  2880   & HR13    & 1.78   & 1.368  \\
 Sep. 02, 2010  &07:41:33.444  &  2880   & HR13    & 1.35   & 1.435  \\
\hline
\end{tabular}
\label{t:logobs}
\end{table}

We used the 1-D, wavelength calibrated spectra as reduced by the dedicated ESO
Giraffe pipeline to derive RVs 
using the {\sc IRAF}\footnote{IRAF is  distributed by the National Optical
Astronomical Observatory, which are operated by the Association of Universities
for Research in Astronomy, under contract with the National Science Foundation }
task {\sc FXCORR}, with appropriate templates, while the RVs of the stars
observed with UVES were derived with the IRAF task RVIDLINES.

Three stars with GIRAFFE spectra and one with UVES spectra
were successively dropped from the sample,
because of unreliable photometry and/or spectra.
Hence our final
sample for NGC~1851 consists in 124 RGB stars, the largest set ever observed
with high resolution spectroscopy in this cluster.  Coordinates, magnitudes and
heliocentric RVs are shown in Tab.~\ref{t:coo18} (the full table is only available in
electronic form at CDS). 

\section{Atmospheric parameters and abundance analysis}

We derived the atmospheric parameters and abundances using the same technique
adopted for the other GCs targeted by our FLAMES survey. In particular, since we
think we have detected a small metallicity spread in NGC~1851, the reference paper 
where the tools are described in detail is the analysis of M~54 (Carretta 
et al. 2010e), although in that case the spread in [Fe/H] is much larger.
Therefore, only a few points of the analysis will be summarised here.

As usual, we derived first estimates of $T_{\rm eff}$\ (from $V-K$ colours)
and bolometric corrections using the calibrations of Alonso et al. (1999, 2001).
Subsequently, we refined the $T_{\rm eff}$\ estimates for the stars using a
relation between $T_{\rm eff}$\ 
and $K$ magnitudes\footnote{For a few stars with no $K$ in 2MASS we used $K$
values obtained by interpolating a quadratic relation as a function of $V$
magnitudes}. Gravities were obtained from stellar masses and radii, these 
last derived from luminosities and temperatures. Within this procedure, the 
reddening and the distance modulus for NGC~1851 were taken from the Harris 
(1996) catalogue (2011 update), and we adopted masses of 0.85~M$_\odot$\ 
for all stars\footnote{Values of surface gravity, derived from the positions 
of stars in the CMD, are not very sensitive to the exact value of the adopted
mass.} and $M_{\rm bol,\odot} = 4.75$ as the bolometric magnitude for the Sun,
as in our previous studies.

The abundance analysis is mainly based on equivalent widths ($EW$s). Those 
measured on GIRAFFE spectra were corrected to the system given by UVES spectra.
Values of the microturbulent velocities $v_t$\ were obtained by eliminating 
trends between Fe~{\sc i} abundances and expected line strength (Magain 1984).
Finally, we interpolated models with the appropriate atmospheric parameters and
whose abundances matched those derived from Fe {\sc i} lines within the Kurucz
(1993) grid of model atmospheres (with the option for overshooting on).
The adopted atmospheric parameters and iron abundances are listed in
Tab.~\ref{t:atmpar18}.

The procedure for error estimates is the same as in Carretta et al. (2010e) and
results are given in the Appendix for UVES and GIRAFFE observations
in Tab.~\ref{t:sensitivityu18} and Tab.~\ref{t:sensitivitym18}, respectively.

As in our previous papers, oxygen abundances are obtained from the forbidden 
[O~{\sc i}] lines at 6300.3 and 6363.8~\AA, after cleaning the former from 
telluric lines as described in Carretta et al. (2007c). The prescription by
Gratton et al. (1999) were adopted to correct the derived Na abundances for
non-LTE effects (from the 5682-88 and 6154-60~\AA\ doublets). Abundances of 
Al (from the 6696-99~\AA\ doublet) and Mg are derived as in Carretta et al.
(2009b). The abundances of these proton capture elements for individual stars
are listed in Tab.~\ref{t:proton18}.

The observed $\alpha-$elements in the present study are Mg (reported among
proton-capture elements), Si, Ca, and Ti. The abundances for these elements
were derived for stars with either UVES or GIRAFFE spectra (see Tab.~\ref{t:alpha18}). 
For the first ones, we measured Ti abundances from both neutral and singly
ionised species, due to the large spectral coverage of UVES spectra\footnote{The
very good agreement obtained for Ti {\sc i} and Ti {\sc ii} in Tab.~\ref{t:meanabu} 
supports our adopted scale of atmospheric parameters, since the ionization balance 
is very sensitive to any detail, and error, in the abundance analysis.}. 

Concerning the Fe-peak elements, we measured lines of Sc~{\sc ii}, V~{\sc i},
Cr~{\sc i}, Co~{\sc i}, and Ni~{\sc i} for stars with GIRAFFE and UVES spectra
and additionally of Mn~{\sc i} and Cu~{\sc i} for stars with UVES spectra. 
Whenever relevant (e.g. Sc, V, Mn, Co), we applied corrections due to the 
hyperfine structure  (see Gratton et al. 2003 for references).
For Cu~{\sc i} we used the transition at 5105~\AA, which we synthesised using a
line list kindly provided by J. Sobeck in advance of publication. The line list
takes into account both isotopic structure and hyperfine splitting.  We adopted
the solar isotopic ratio of 69\% $^{63}$Cu and 31\% $^{65}$Cu.
Abundances of Fe-peak elements for individual stars are reported in
Tab.~\ref{t:fegroup18}.

We derived the concentration of several neutron capture elements, mostly from UVES 
spectra. We measured the abundances of Y~{\sc ii}, Zr~{\sc ii}, Ba~{\sc ii}, La~{\sc ii},
Ce~{\sc ii}, Nd~{\sc ii}, Eu~{\sc ii}, and Dy~{\sc ii} through a combination of 
spectral synthesis and EW measurements. A few details follow for each element.

\begin{figure*}
\centering
\includegraphics[scale=0.7]{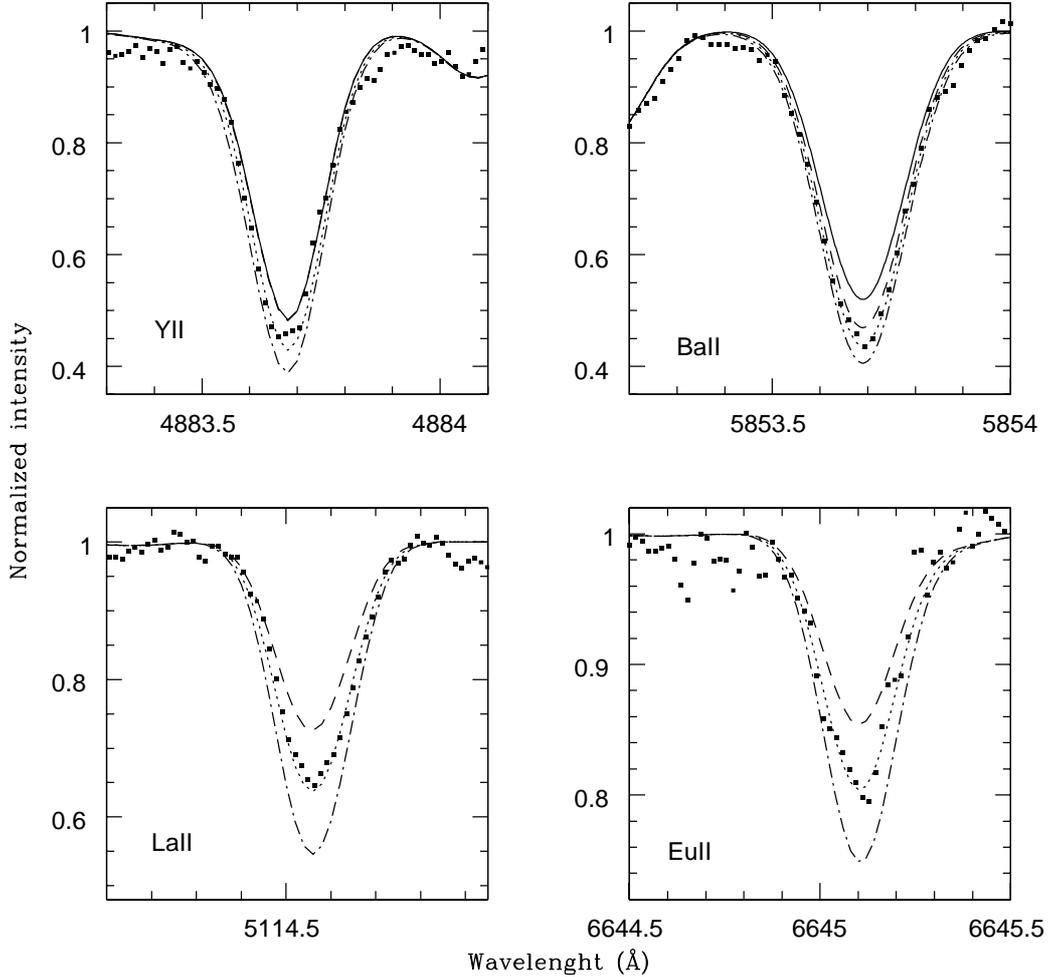}
\caption{Examples of syntheses for lines due to four $n-$capture species in star
14080. Upper left side: syntheses for the Y~{\sc ii} line at 4883\,\AA, the
abundances plotted are [Y/Fe]{\sc ii}=0.1, 0.3 and 0.5\,dex.  Upper right side:
as upper left for the Ba~{\sc ii} line at 5853.5\,\AA\, for [Ba/Fe]{\sc ii}=0.30,0.50 and
0.70\,dex. Lower right side: as upper left for the Eu line at 6645.1\,\AA\, for
[Eu/Fe]{\sc ii}=0.4, 0.6,0.8; lower left side:  as for upper left for the
La~{\sc ii}
line 5114.5\,\AA\,  for [La/Fe]{\sc ii}=0.2, 0.4 and 0.6\,
dex. For Y and Ba the solid lines indicate the values reported in V10 for this star.}
\label{f:synt}
\end{figure*}

\paragraph{Y~{\sc ii}:} We used four transitions (at 4883, 5085, 5200, 5206~\AA). 
They were synthesised using the same atomic parameters as listed in Sneden 
et al. (2003). An example of the fit of the 4883~\AA\ line for star 14080 
is given in Fig. \ref{f:synt}.

\paragraph{Zr~{\sc ii}:} The line at 5112~\AA\ is the only clean usable line 
included in our spectral range. We adopted the $gf$\ and excitation potential
listed by Biemont et al. (1981).

\paragraph{Ba~{\sc ii}:} Four lines were used to estimate the abundance: 4934.1, 
5853.7, 6141.7, and 6496.9~\AA. We used spectral synthesis for the bluest line, 
since it is affected by considerable blending with 
a strong Fe line in the spectra under consideration. For the remaining lines, the abundance was measured through 
their EWs, because they are clean from blending for the stars of NGC~1851.
The transition parameters adopted are from Sneden et al. (1996). An example of
the fit of the 5853~\AA\ line for star 14080 is given in Fig. \ref{f:synt}.
Even though we derived the abundance from EWs for this particular line, we
chose
it to display a synthesis to show that the Ba abundance from V10 for this particular 
star is not a good fit to our data (at least using our adopted atmospheric parameters).
We note that abundances from synthesis and EW agree very well for the plotted line.

\paragraph{La~{\sc ii}:} We used four lines (4921.0, 4921.8, 5114.5 and 6262.3\,\AA).
The three bluest ones were synthesized, as they have a not negligible hyperfine
structure. Abundances from the other line were derived from EWs measurements. The
transition parameters adopted are from Sneden et al. (2003). An example of the
fit of the 5114.5~\AA\, line for star 14080 is given in Fig. \ref{f:synt}.

\paragraph{Ce~{\sc ii}:} We used one line at 5274.3\,\AA. The transition parameters
adopted are from Sneden et al. (2003). We assign a smaller weight to our Ce abundances,
because of possible blending of this line.
   
\paragraph{Nd~{\sc ii}:} We used 9 lines for which we measured EWs  (5092.8, 5130.6,
5165.1, 5212.4, 5234.2, 5249.6, 5293.2, 5311.5, and 5319.8~\AA). The transition
parameters adopted are from Sneden et al. (2003). 

\paragraph{Eu~{\sc ii}:} We synthesised two spectral lines, at 6437.6 and 6645.1~\AA.
This technique was used because they are
affected by considerable hyperfine splitting. The transition parameters adopted 
are from Sneden et al. (2003). An example of the fit of the 6645~\AA\ line for
star 14080 is given  in Fig. \ref{f:synt}.

\paragraph{Dy~{\sc ii}:} Abundance was derived from EW of the line at 5169.7\,\AA. The
transition parameters adopted are from Sneden et al. (2003). 

Results are given in
Tab.~\ref{t:neutron18} and Tab.~\ref{t:zrba18} for stars with UVES and GIRAFFE
spectra, respectively.
The averages of all measured elements with their $r.m.s.$ scatter are listed in 
Tab.~\ref{t:meanabu}.

\setcounter{table}{8}
\begin{table}
\centering
\caption{Mean abundances from UVES and GIRAFFE}
\begin{tabular}{lcc}
\hline
                     &               &               \\
Element              & UVES	     & GIRAFFE       \\
                     &n~~   avg~~  $rms$ &n~~	avg~~  $rms$ \\        
\hline
$[$O/Fe$]${\sc i}    &13   +0.04 0.25 &~86 $-$0.01 0.21\\
$[$Na/Fe$]${\sc i}   &13   +0.26 0.25 &120   +0.19 0.28\\
$[$Mg/Fe$]${\sc i}   &13   +0.35 0.03 &119   +0.38 0.04\\
$[$Al/Fe$]${\sc i}   &13   +0.13 0.21 &  	       \\
$[$Si/Fe$]${\sc i}   &13   +0.38 0.03 &120   +0.39 0.03\\
$[$Ca/Fe$]${\sc i}   &13   +0.30 0.02 &121   +0.33 0.04\\
$[$Sc/Fe$]${\sc ii}  &13 $-$0.01 0.08 &121   +0.03 0.06\\
$[$Ti/Fe$]${\sc i}   &13   +0.17 0.07 &118   +0.15 0.05\\
$[$Ti/Fe$]${\sc ii}  &13   +0.17 0.05 &  	       \\
$[$V/Fe$]${\sc i}    &13 $-$0.06 0.09 &116 $-$0.13 0.12\\
$[$Cr/Fe$]${\sc i}   &13 $-$0.02 0.06 &113   +0.05 0.11\\
$[$Mn/Fe$]${\sc i}   &13 $-$0.34 0.08 &    	       \\
$[$Fe/H$]${\sc i}    &13 $-$1.18 0.07 &121 $-$1.16 0.05\\
$[$Fe/H$]${\sc ii}   &13 $-$1.14 0.12 &101 $-$1.13 0.07\\
$[$Co/Fe$]${\sc i}   &13 $-$0.24 0.06 &~49 $-$0.03 0.07\\
$[$Ni/Fe$]${\sc i}   &13 $-$0.10 0.03 &120   +0.02 0.07\\
$[$Cu/Fe$]${\sc i}   &13 $-$0.46 0.07 &                \\
$[$Y/Fe$]${\sc ii}   &13   +0.27 0.15 &  	       \\
$[$Zr/Fe$]${\sc ii}  &13   +0.26 0.11 &  	      \\
$[$Ba/Fe$]${\sc ii}  &13   +0.48 0.26 &101   +0.49 0.22\\
$[$La/Fe$]${\sc ii}  &13   +0.23 0.17 &  	     \\
$[$Ce/Fe$]${\sc ii}  &13   +0.69 0.20 &  	     \\
$[$Nd/Fe$]${\sc ii}  &12   +0.67 0.15 &  	     \\
$[$Eu/Fe$]${\sc ii}  &13   +0.67 0.11 &  	     \\
$[$Dy/Fe$]${\sc ii}  &13   +0.67 0.12 &  	     \\
\hline
\end{tabular}
\label{t:meanabu}
\end{table}

\section{Metallicity spread and stellar population components}

\subsection{Metallicity spread in NGC~1851}

\begin{figure}
\centering
\includegraphics[scale=0.44]{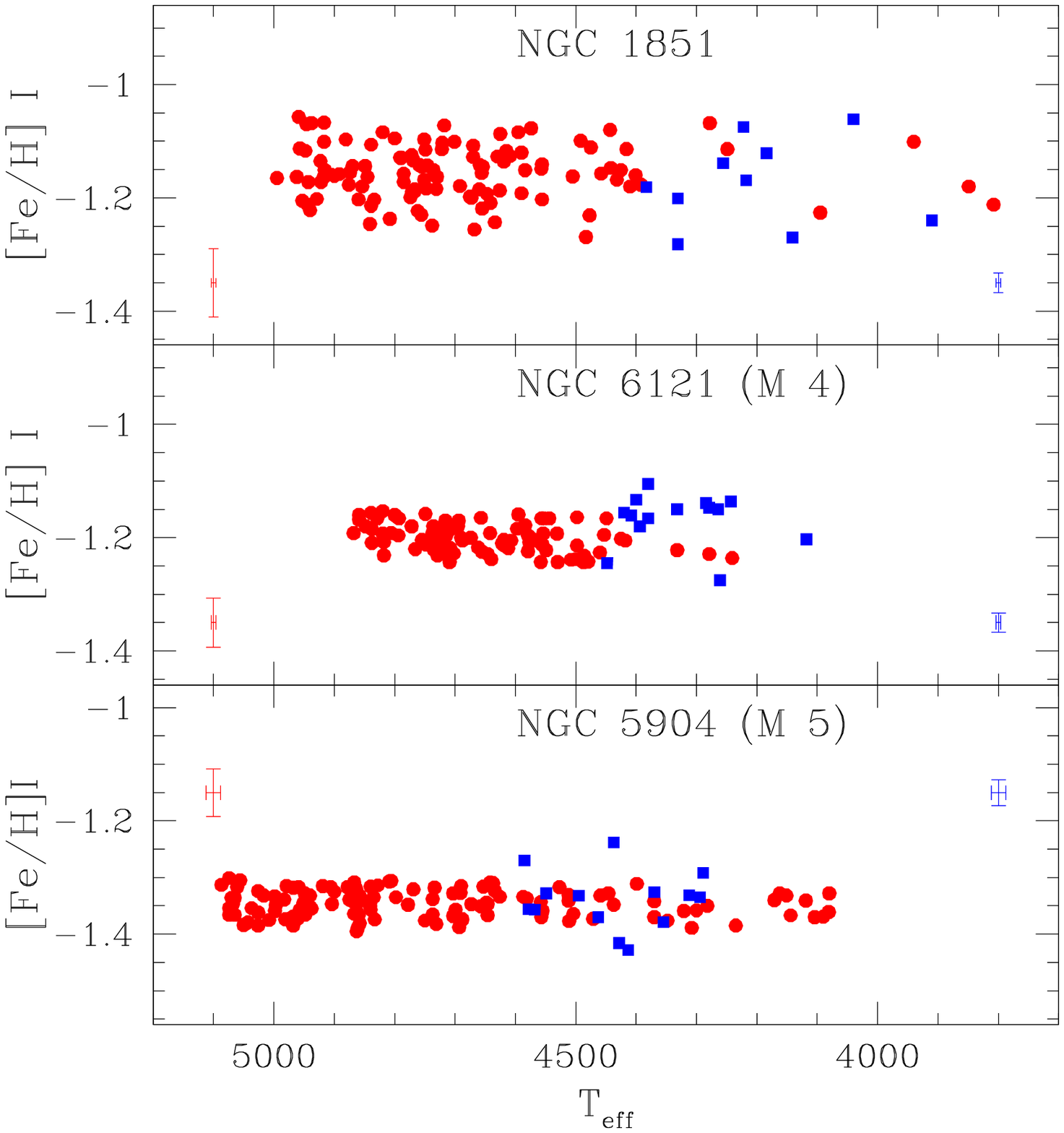}
\caption{[Fe/H] ratios as a function of effective temperature for stars analysed
in NGC~1851 (upper panel), M~4 (middle panel) and M~5 (lower panel),
both from Carretta et al. 2009a,b). The filled red circles indicate stars with
GIRAFFE spectra, the blue filled squares stars with UVES spectra. 
Error bars in the left and right side of the panels refer
respectively to our GIRAFFE and UVES data and indicate star-to-star errors.}
\label{f:feteff185961}
\end{figure}

In Carretta et al. (2010c) we argued we detected a small but real metallicity 
spread in NGC~1851 by comparing observed abundances with internal errors. 
We deem that result robust because it rested on a statistically significant 
sample. NGC~1851 shows a much larger dispersion in [Fe/H] values when compared to 
other GCs of similar metal abundance, examined with the same instrumentation 
and technique of analysis (M4 and M5: see Fig.~\ref{f:feteff185961}).

\begin{figure}
\centering
\includegraphics[scale=0.42]{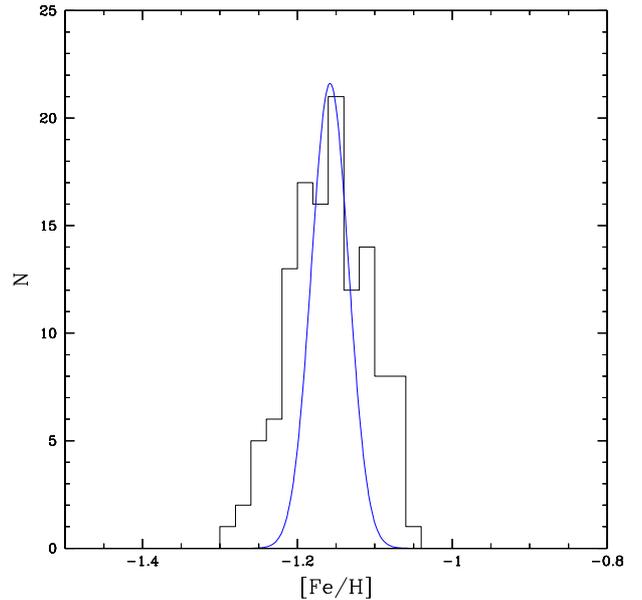}
\caption{Histogram of [Fe/H] ratios for our NGC~1851 stars. For the gaussian
curve we adopted as FWHM=0.024, that is the average standard deviation in [Fe/H]
between the mono-metallic GCs M~4 (rms=0.025) and M~5 (rms=0.023, see Carretta et
al. 2009c).}
\label{f:gaussian}
\end{figure}

Apart from a larger scatter among the brighter stars observed with UVES
(discussed in detail in Carretta et al. 2009c), the iron dispersion observed in
M~4 and M~5 is compatible with internal errors associated to the analysis, 0.025
and 0.023 dex, respectively. However, if we try to fit the iron distribution in
NGC~1851 using a Gaussian with FWHM equal to 0.024 dex (the average error from
M~4 and M~5, Fig.~\ref{f:gaussian}) the plot clearly indicates the presence of a
small intrinsic dispersion in the iron content that cannot be accounted for by
internal errors alone.

On the other hand, the comparison of the scatter in Fe abundances 
(Tab.~\ref{t:meanabu}) with the internal errors estimated in the Appendix
shows that the observed dispersion in [Fe/H]~{\sc i} for NGC~1851 are
statistically significant for stars with metallicity from UVES,
as pointed out in Carretta et al. (2010c). In this case, internal errors
are of the order of 0.03 dex, much lower than the observed scatter (0.07~dex).
The case is less compelling for the abundances from GIRAFFE data, for which, 
we obtained an internal error on [Fe/H] (0.043~dex) not much lower 
than the observed scatter ($0.051\pm 0.005$~dex). 
However, putting together the highly significant result obtained from UVES
spectra with the hint coming from GIRAFFE ones and the consideration
that the observed scatter is much larger than for the similar clusters M4 
and M5, we conclude in favour of a real dispersion in [Fe/H] values within
NGC~1851. 

A similar amount of intrinsic scatter was also found by Yong and Grundahl
(2008). The eight stars observed in that study have an average [Fe/H]=$-1.268\pm
0.030$ with a 1$\sigma$ r.m.s. scatter of 0.084 dex. Although there is a small
offset with respect to the present analysis (likely  due to different
temperature and $gf$\ scales adopted), their study agrees with us on the reality
of a small metallicity dispersion in NGC~1851. On the other hand, V10
did not find any scatter in [Fe/H] among their 15 stars.

Given the method used to derive atmospheric parameters for individual stars, a 
potential contamination by AGB stars in our sample could increase the observed
scatter. This might occur in particular for the UVES sample, where most of the stars are brighter
than the magnitude limit at which it is easy to discriminate AGB and RGB stars
(see Fig.~\ref{f:cmdsel1851}). From this diagram the AGB is clearly discernible
from the RGB below $V\sim 14.5$ ($T_{\rm eff}\sim 4300$~K and $M_V\sim -1$). For fainter
stars, we simply omitted AGB stars from our sample. For brighter stars, the two 
sequences do not appear any more distinct, because such bright AGB stars are 
$< 50$~K warmer than first ascent RGB stars (this value is derived from the 
isochrone by Bertelli et al. 2008 at Z=0.001 and 12.6 Gyr). Our sample may then 
include some bright AGB stars. However, we believe that this is not a source of concern 
owing to several reasons. First, we expect very few AGB stars, since they are less 
than about 20\% of the red giants: hence on 17 observed stars brighter than the limit 
we expect only about three AGB interlopers. Second, the errors produced in the analysis 
are small. In fact, even if we make an error of 0.2~$M_\odot$, by mistaking AGB stars 
for RGB stars, this translates into an error of 0.1~dex in $\log{g}$, all the other 
parameters being held fixed; and, as mentioned above, the errors made in the $T_{\rm eff}$\ 
values are $< 50$~K. Using the sensitivities listed in Tab.~\ref{t:sensitivitym18} and 
Tab.~\ref{t:sensitivityu18} in the Appendix, the gravity error would imply errors 
in the derived [Fe/H] ratio of $-0.003$ dex and $+0.012$ dex for GIRAFFE and UVES, 
respectively. Due to errors in temperature, bright AGBs confused in the RGB 
sample could appear up to be $0.04 \div 0.06$ dex more metal-poor than the other stars.
These effects are very small. Furthermore, we notice that the observed
metallicity scatter is about
constant over the whole temperature range, either including or not the temperature
range where possible contamination by AGB stars exist (see Fig.~\ref{f:feteff185961}).

Therefore, the observed spread seems real, and not due to a neglected
contamination from AGB stars. This is corroborated by the good agreement we obtain,
on average, between abundances based on neutral and singly ionised species (e.g.
iron) that supports the atmospheric parameters we used, including surface 
gravities $\log{g}$.  More in general, we do not attribute much weight to the
small offset between the abundances of neutral and singly ionised iron, which is
well within 1$\sigma$ r.m.s. scatter, thus hardly significant.  The [Fe/H]~{\sc ii}
abundances rest on only one-two lines for GIRAFFE spectra and bright stars
(those observed with UVES) are also cooler, on average, hence more subject to
possible NLTE effects. Anyway, we do not use [Fe/H]~{\sc ii} in any of our
conclusions.

Finally, we stress that in the present analysis the observed spreads are only
{\it lower} limits, owing to the particular way we derive the final temperatures,
using a {\it mean} relation between T$_{\rm eff}(V-K)$\ and the $K$\ magnitude along the RGB.
While this technique allows us to have extremely small internal errors (see
Carretta et al. 2007a for details), it also tends to slightly reduce any real 
metallicity spread along the RGB, leading us to adopt an effective temperature 
corresponding to a mean ridge line. Had we iteratively used slightly
different temperature-magnitude relations for each component, we would have 
obtained not very different results, on average, the maximum additional
difference in [Fe/H] amounting to less than 0.04 dex.

Anyway, on top of these results, the most striking evidence of a difference in
the stellar populations on the RGB in NGC~1851 is the spatial distribution of
the metal-rich (MR) and  metal-poor (MP) components (Carretta et al. 2010c). We
found there a clearcut evidence that stars more metal-poor than the cluster
average ([Fe/H]$=-1.16$) are more centrally concentrated than more metal-rich
stars, with a very high degree of confidence. {\it There is something
intrinsically different between the MR and MP populations in NGC~1851}.

This finding bypasses any debate on the reality of radial distribution of SGB
stars in the existing literature and was possible only owing to the large number
of stars analysed in a very homogeneous way. This observation stimulated the
working hypothesis proposed in Carretta et al. (2010c) that NGC~1851 might be
the result of a merger of two originally distinct clusters, with a slightly
different metallicity.

There are other differences between the MR and MP components, beside [Fe/H] and
radial distribution, in the content of $\alpha-$capture and neutron-capture
elements (see below). Moreover, in each individual component there is a well
developed Na-O anticorrelation (Carretta et al. 2010c), which is the chief
signature of a genuine globular cluster, according to Carretta et al. (2010a).
Very recently, a very similar finding has been obtained for M~22 (Marino et al.
2011), which has several analogies with NGC~1851.

\subsection{A cluster analysis approach to stellar components in NGC~1851}

While the choice of the average [Fe/H] as separating value between the two
populations is the simplest one, it is also arbitrary. To seek for a
more objective criterion we exploit the main results obtained up to now for
NGC~1851 (Carretta et al. 2010c): (i) there is a real (small) metallicity
dispersion in the cluster, where (ii) the MP component is more centrally
concentrated, and moreover (iii) the MR component shows a higher level of
$s-$process elements. These observations suggest to use the run of [Fe/H] 
as a function of [Ba/H] as a population diagnostic to separate the two stellar 
components in NGC~1851. On this data, we performed a cluster analysis using
the $k-$means algorithm (Steinhaus 1956; MacQueen 1967), as implemented
in the $R$\ statistical package (R Development Core Team, 2011), a system 
for statistical computation and graphics, freely available 
on-line\footnote{http://www.R-project.org}. Two key features of $k-$means are:
\begin{itemize}
\item Euclidean distance is used as a metric and variance is used as a measure of 
cluster scatter. The various parameters used should be transformed before computing
these distances, in order to weight them adequately. A sensible choice is to normalize
them to the observational errors. In practice, as normalization factors we assumed
errors of 0.04 and 0.20 dex for [Fe/H] and [Ba/H] respectively.
\item The number of clusters $k$\ is an input parameter: an inappropriate choice of $k$\ 
may yield poor results. Given that the ratio between observed scatter of data and
observational errors is not large, we adopted $k=2$. Of course, there may be more
than two groups of stars in NGC~1851, but we wished to limit the number of hypothesis
at this stage.
\end{itemize}
A key limitation of $k-$means is its cluster model. The concept is based on spherical
clusters that are separable in a way so that the mean value converges towards the cluster 
center. The clusters are expected to be of similar population, so that the assignment to 
the nearest cluster center is the correct assignment. When the clusters have very
different size, this may result in poor assignation of members to clusters. However,
in the present case, we expect that most of the scatter within one cluster is due to
observational errors, which are similar for the different clusters. We then 
considered the assumption of similar size for the different clusters acceptable.
Of course, this does not mean that occasionally assignation of some member (that
is, star) to a particular cluster is questionable.

\begin{figure}
\centering
\includegraphics[scale=0.44]{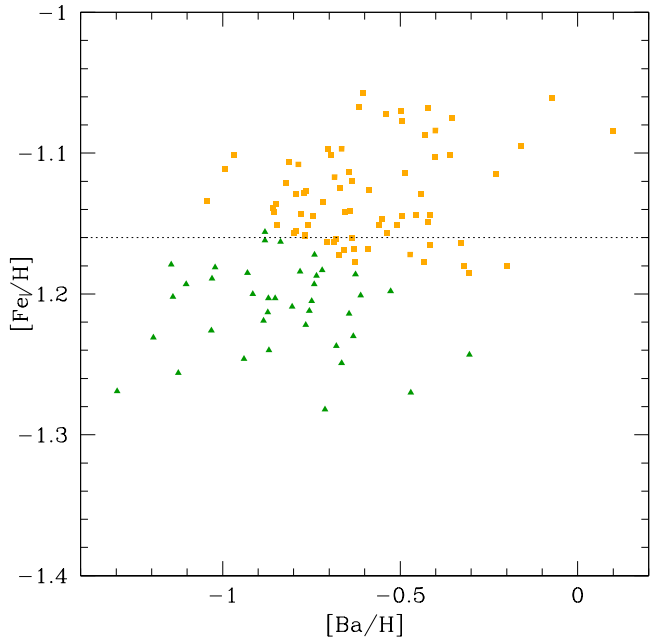}
\caption{Diagnostic diagram [Fe/H] vs [Ba/H] for the stellar components in
NGC~1851. The different symbols and colour coding show the results of the
cluster analysis using Fe and Ba abundance as parameters (see text).
Stars indicated with orange filled squares are attributed to the MR component;
stars indicated with green filled triangles are assigned to the MP component.
The dotted line indicates the average value [Fe/H]$=-1.16$ dex for
NGC~1851 which we previously adopted as separating value.}
\label{f:fehbah}
\end{figure}

The results are shown in Fig.~\ref{f:fehbah}, where the two components are
indicated with different symbols and colours. We also superimpose a line
indicating the average value [Fe/H]$=-1.16$ dex for NGC~1851, adopted by
Carretta et al. (2010c) to separate the two components using only the
metallicity. 

Notice that 13 stars did not have Ba abundances because lacking of UVES or HR13
GIRAFFE spectra. For 10 stars it is easy to unambiguously assign a 
population simply according to the metallicity (most of them falling above
[Fe/H]$=-1.10$ or below [Fe/H]$=-1.20$ dex). However, for three stars 
(28116, 33385, 40300) this is not possible and their status is currently
uncertain.

The cluster analysis shows that the new components are similar to, but not 
exactly coincident with, those defined in Carretta et al. (2010c). The separation 
line is slightly slanted and not simply an average value, when accounting also 
for the enhanced level of Ba in the most metal-rich component. However, for 
simplicity, we continue to talk of MR and MP components.
The main change is that the MP component decreases from 50\% to 36\% while the
MR one increases to 64\%. These fractions are now tantalisingly close to the
fractions of stars observed on one hand on the faint SGB (45\%) and the BHB
(40\%), and on the other hand on the bright SGB (55\%) and RHB (60\%, see
Milone et al. 2008). This might suggest a tentative link according to the new
population ratios, although this link should actually be confirmed by
determination of abundances for SGB and HB stars.

\begin{figure}
\centering
\includegraphics[scale=0.42]{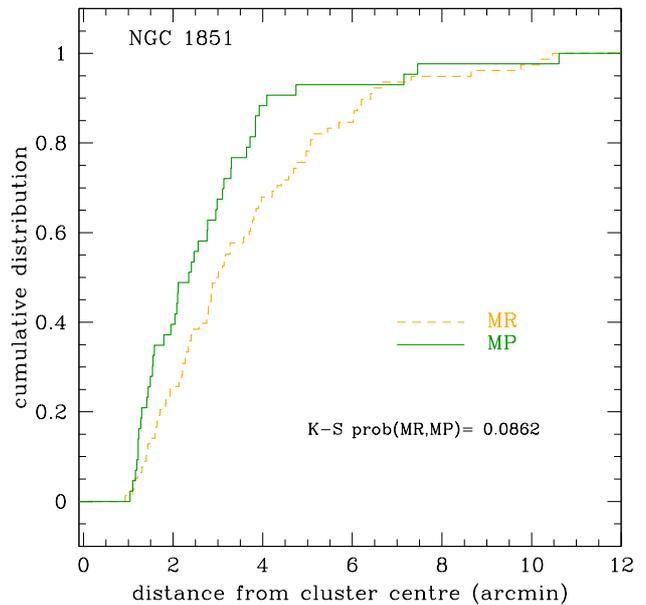}
\caption{Cumulative distribution of radial distances for all stars in our sample, divided
in MP (green solid line) and MR (dashed orange line) components.}
\label{f:distrFe}
\end{figure}

Also with this new definition the MP component results to be more
concentrated than the MR one to a reasonable level of confidence 
(see Fig.~\ref{f:distrFe}). 

Milone et al. (2009) considered the issue of the radial distribution of the
different populations in NGC~1851, claiming no difference between the two
SGBs. Their sample is much more numerous than ours,
and since their conclusion is apparently different,
we re-examined their result. In their analysis, they considered three possible
cases (labelled A, B and C), depending on the interpretation of the splitting of the SGB.
Case A is for a simple age difference (about 1 Gyr); case B is for a combination
of age differences (about 2 Gyr) and CNO abundance variations; and case C is
for a difference only in the sum of CNO abundances. If we use as metrics for the
comparison the Pearson correlation coefficient between the radius and the fraction
of faint SGB stars, there is slightly more than 5\% of probability that
the correlation is random for cases A and B, while there is no clear difference for
case C. Since they seems to prefer this last scenario, Milone et al. (2009) concluded
that there is no difference in the radial distribution of the different
populations of NGC~1851. However, much more could be said apropos.
First, obviously lack of evidence is not evidence of lack. Second, in their data
there is indeed some hint for a radial gradient, the faint SGB being more concentrated
than the bright SGB. While the test considered above seems not strong enough to allow a
conclusion at a high confidence level, other tests like the Kolmogorov-Smirnov one
are more powerful in revealing differences among distributions,
and would seem more appropriate in this context. While we have not at our disposal 
enough data to perform the Kolmogorov-Smirnov test on their data, we might use other
metrics. For instance, we may compare the weighted averages of the
three innermost and of the three outermost bins. Using this approach, we found
differences of $0.11\pm 0.06$, $0.17\pm 0.07$, and $0.06\pm 0.06$\ for cases A, B, and C.
Summarizing, the lack of evidence claimed by Milone et al. (2009) is related to the 
statistics they used in their analysis and to the interpretative scenario
they favour for the separation of NGC~1851 into two populations. The statistics used
is likely not the most powerful one, and there is no strong spectroscopic evidence
for a variation of the sum of CNO abundances in NGC~1851 (see also V10).
We then think that their data cannot be used to conclude that there is no difference
in the radial distributions of the different populations of NGC~1851.

\subsection{Kinematics of stellar components in NGC~1851}

\begin{figure}
\centering
\includegraphics[scale=0.40]{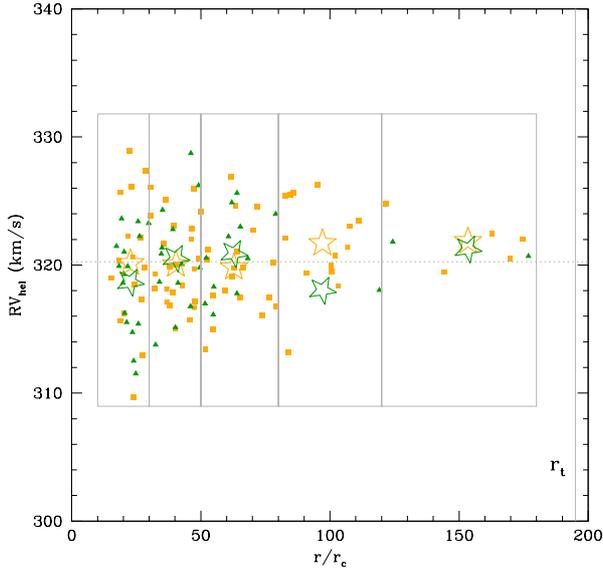}
\caption{Distribution of radial velocities for all stars in our sample, divided
in MP (filled greed triangles) and MR (filled orange squares) components. The x
axis shows the distance from the cluster centre (in units of the core radius 
$r_c$). The larger symbols, also in green and orange, represent the average RV
of the two sub-samples, computed in the boxes indicated in the figure. The
dotted horizontal line represents the global average of the sample (320.26
km~s$^{-1}$, with a rms of 3.74) and the vertical line indicates the tidal
radius. The average RVs at the various distances are given in 
Table~\ref{t:rvdist}.}
\label{f:newfigvr}
\end{figure}

In principle, if the MP and MR components come from two distinct sub-units they
could also have different kinematics, showing up in their velocity distribution.
However, we found no significant differences: the
velocity dispersion increases from the outskirts towards the cluster centre (see
Fig.~\ref{f:newfigvr} and Tab.~\ref{t:rvdist}) but for the MR and MP the pattern
is similar, apart from the different concentration of stars in the two
components. 

\begin{table}
\centering
\caption{Average RVs for MP and MR stars at different distances from the cluster
centre}
\begin{tabular}{rccrccr}
\hline\hline
Dist.  &$<RV(MR)>$   &rms &Nr.     &$<RV(MP)>$     &rms &Nr. \\
(in $r_c$) &(km s$^{-1})$ &    &MR        &(km s$^{-1}$) &    &MP  \\
\hline
 10-30     &320.16      &5.25&16      &318.67       &3.90&16 \\
 30-50     &320.05      &3.26&23      &320.55       &4.25&13 \\
 50-80     &319.82      &3.51&20      &320.91       &3.30&11 \\
 80-120    &321.70      &3.59&14      &318.04       &0.00& 1 \\
 $>$120     &321.86      &2.05& 5      &321.25       &0.78& 2 \\
\hline
\end{tabular}
\label{t:rvdist}
\end{table}

Given the limitations of our sample, we cannot tell if this is due 
to insufficient data or it has some real physical meaning. We only
notice that if the merger is recent, there could be only a small
dynamical effect, because after the merging the cluster would not be relaxed,
and there might be no energy equipartition between the different populations. 
The merging between the two putative clusters could well be an event occurred 
only a few Gyrs ago: the relaxation time at the half mass radius (and we
sampled a region external to this radius)
is about 0.7 Gyr (Harris 1996), and a 
full cluster relaxation does require some 2-3 relaxation times.
In the framework outlined in Carretta et al. (2010a) for the formation model 
of globular clusters, the only constraint is that the merger between the two
clusters must have occurred before their host dSph galaxy merged with our
Galaxy. However, we do not know when this possible interaction happened. If 
some debris of this putative dSph still exist (as suggested by several authors, 
e.g. Carballo-Bello et al. 2010, Olszewski et al. 2009) then the event could be
rather recent. More data and detailed dynamical modelling would be required;
however, this is beyond the purpose of the present work, which is focused on
the chemical tagging.

\section{The complex nucleosynthesis in NGC~1851}

\subsection{Nucleosynthesis in cluster environment: the proton capture elements}

The pattern of observed (anti)correlations among light elements like C, N, O, F,
Na, Mg, Al, Si requires a specific chain of events, restricted to the
proto-cluster environment (see e.g. Gratton et al. 2004). Unavoidable
requirement for a GC is the presence of a first generation of stars providing at
their death the material, processed through proton-capture reaction in H-burning
at high temperature, that participated to the formation of the second stellar
generation(s) with modified chemical composition. 

These considerations, together with the existence of the Na-O anticorrelation 
among stars of {\it each} metallicity component on the RGB of NGC~1851, support
the merger of two clusters as a viable origin to explain the observed characteristics 
of this cluster. We will now examine whether this simple working hypothesis is 
robust against our revised classification (Sect. 4.2) of stellar populations 
and the full set of abundances derived for NGC~1851.

\begin{figure}
\centering
\includegraphics[scale=0.44]{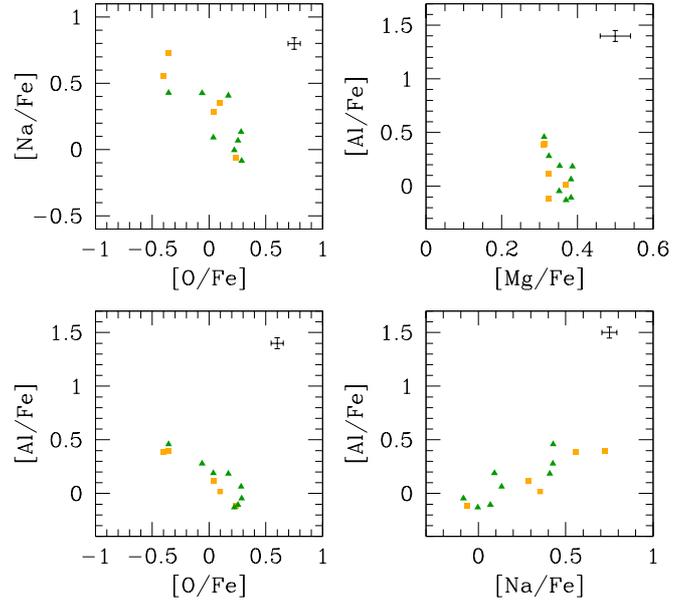}
\caption{Na-O (top-left panel), Mg-Al (top-right panel), Al-O (bottom-left
panel) anticorrelation, and Na-Al correlation (bottom-right panel) for stars with
UVES spectra in NGC~1851. Green triangles and orange squares indicate stars of
the MP and MR component, respectively. Error bars are star-to-star errors.}
\label{f:light4u18}
\end{figure}

The (anti)correlations among proton-capture elements in NGC~1851 are
summarised in Fig.~\ref{f:light4u18} for stars with UVES spectra, from 
which the whole set of Na, O, Mg, Al abundances is available.
NGC~1851 shows a moderate Na-O anticorrelation, and a well defined Na-Al
correlation, with a not extremely high production of Al.
In Fig.~\ref{f:light4u18} there is a hint that stars of the MR component
may reach more extreme processing than stars of the MP component, 
but this sample is limited.

\begin{figure*}
\centering
\includegraphics[bb=18 142 590 477, clip, scale=0.52]{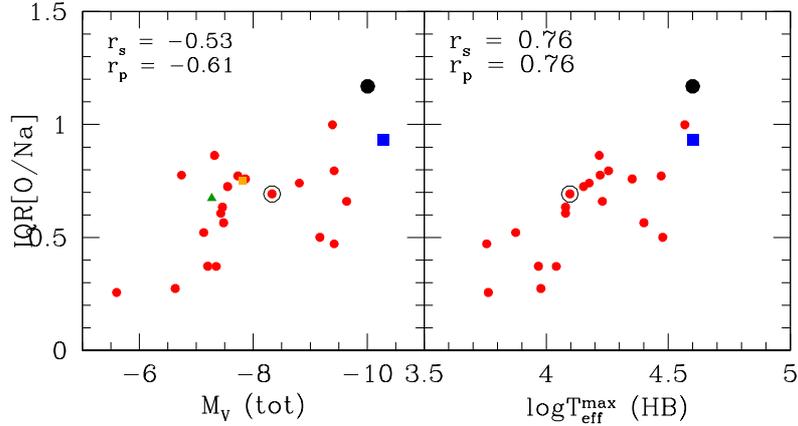}
\caption{Interquartile range of the [O/Na] ratio, IQR[O/Na], for the stars in
NGC~1851 as a function of the total absolute visual magnitude (left panel) and
of the maximum temperature reached by HB stars (mainly from Recio-Blanco et al. 2006,
right panel). NGC~1851 is indicated by a circled filled point.  The large black
filled circle and blue square are for M~54 and $\omega$ Cen, respectively.
The IQR values come from Carretta et al. (2010b) and from data by Johnson and
Pilachowski (2010), respectively, and maximum temperatures estimated by
Cassisi (2011, private communication). The Spearman rank correlation coefficient
($r_s$) and the linear regression Pearson's correlation  coefficient ($r_p$)
are listed in each panel. In the left panel, the green triangle and the orange
square indicate the values relative to the MP and MR components in NGC~1851 (see
text).}
\label{f:iqrmvteff}
\end{figure*}

The global Na-O anticorrelation presents stars well mixed over all the 
metallicity range, and its extension (as measured by the interquartile 
range IQR[O/Na]=0.693) well fits the strong correlations with the total cluster 
mass (using the absolute visual magnitude $M_V=-8.33$ as a proxy) and with the 
maximum temperature along the HB derived and widely discussed in  Carretta et
al.(2010a) (see Fig.~\ref{f:iqrmvteff})\footnote{Note that the analysis 
considered here for $\omega$~Cen (Johnson and Pilachowski 2010) is not 
strictly homogeneous with the other determinations showed in the plot,
which come from our FLAMES survey.}.

\begin{figure*}
\centering
\includegraphics[bb=18 409 590 710, clip, scale=0.52]{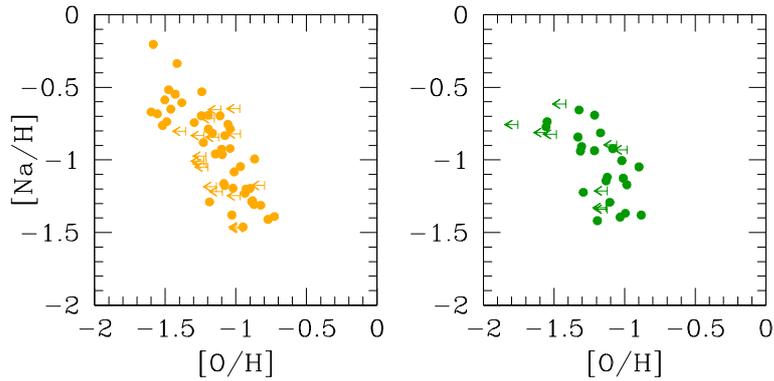}
\caption{Na-O anticorrelation for stars of the MR (left panel) and of the MP 
(right panel) component in NGC~1851.}
\label{f:m18antiohseq12}
\end{figure*}

The cluster analysis, based only on Fe and Ba abundances, separates stars 
of NGC~1851 in two sub-groups, each one presenting a clean Na-O anticorrelation. 
The anticorrelation is slightly more extended for the MR component (IQR[O/Na]=0.750) 
than for the MP one (IQR[O/Na]=0.674) (see Fig.~\ref{f:m18antiohseq12}). 
As an exercise, we tentatively distributed the total $M_V=-8.33$ of
NGC~1851 over the two MR and MP components by using the number of the RHB (191) 
and of BHB stars (116), respectively, adopted by Gratton et al. (2010b) in their
study of the HB in this and other GCs. The resulting values are $M_V=-7.81$ and
$M_V=-7.27$ for the MR and MP groups, respectively. When coupled with the
IQR values derived above, the two components of NGC~1851 nicely fit the overall 
relation between $M_V$ and IQR[O/Na] (green and orange symbols in the left 
panel of Fig.~\ref{f:iqrmvteff}). This supports the idea that the two components 
in NGC~1851, selected using independent parameters (Fe, Ba), do behave as two
individual GCs with slightly different total masses.

\begin{figure}
\centering
\includegraphics[scale=0.42]{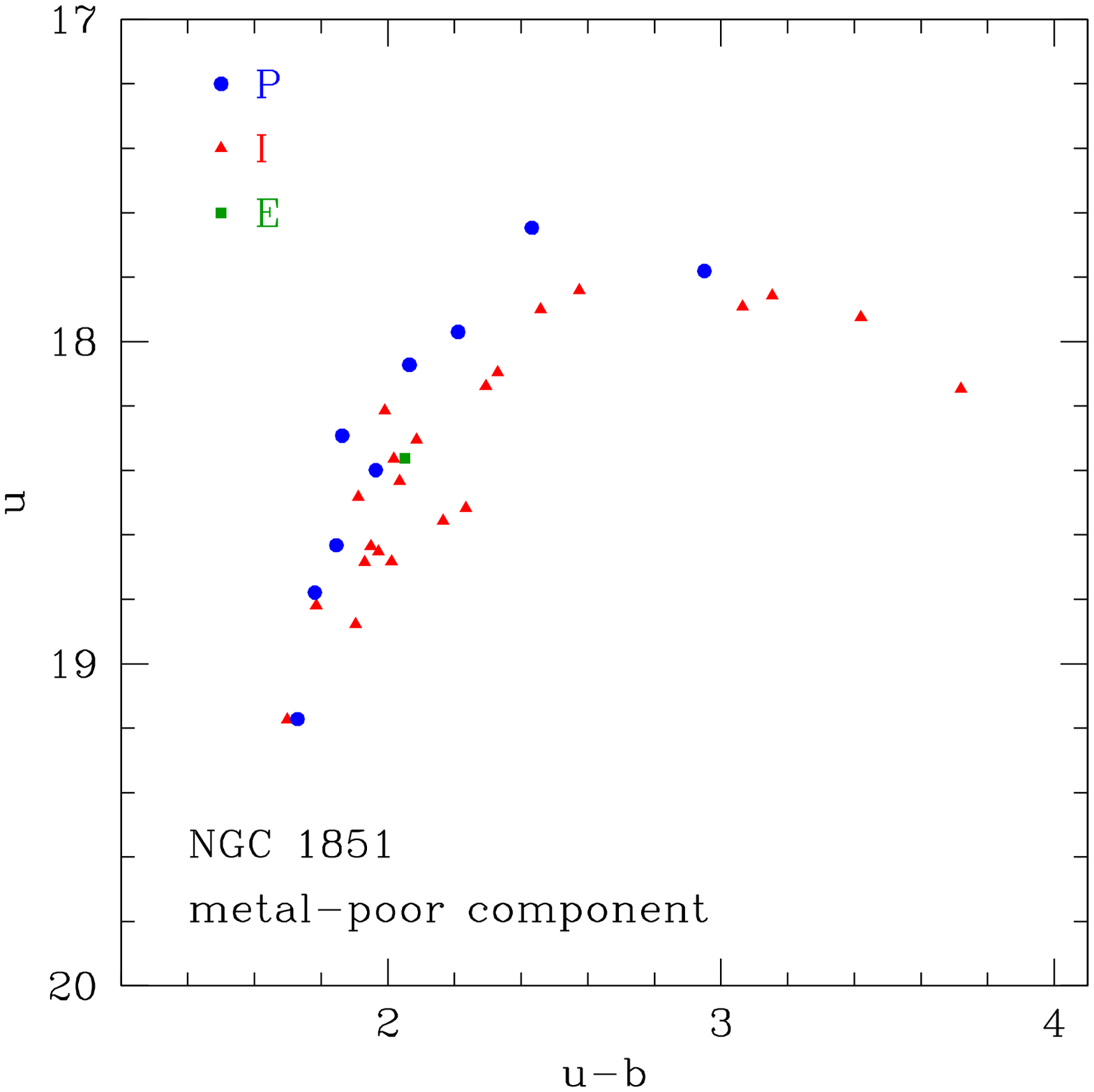}
\includegraphics[scale=0.42]{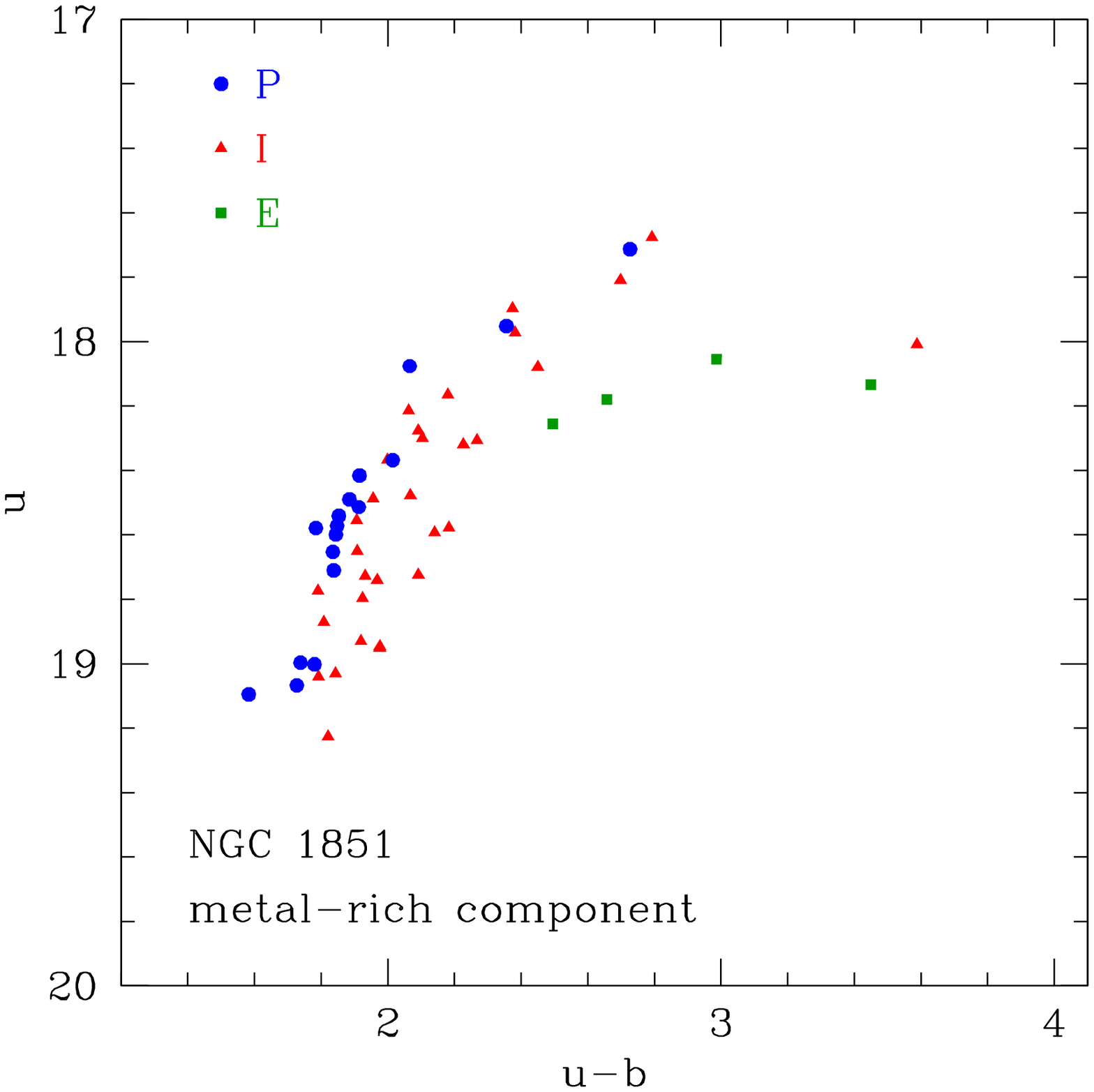}
\caption{Str\"omgren $u,u-b$ CMD for stars of the MP (upper panel) and
of the MR component (lower panel) in NGC~1851. In each panel, blue filled circles
represent the P (first generation) group, the red triangles the intermediate I
group, and the green squares the extreme E group of second generation stars.}
\label{f:ubuPIE}
\end{figure}

To further support this evidence, we used the Str\"omgren photometry by
Grundahl (available through the database used by Calamida et al. 2007) to plot
the stars of the P, I and E components in NGC~1851. 
The definition of the primordial (P), intermediate (I) and extreme (E)
population in a cluster, based on Na and O abundances, is illustrated in
Carretta et al. (2009a), where we showed how the P (Na-poor, O-rich) stars of
the first generation lie on a narrow strip to the blue of the RGB, while the I+E
second generation stars are spread out to the red in a Str\"omgren $u,u-b$ CMD.
The $u,u-b$ CMD is a good plane where to look for this 
segregation, since the splitting is due to the effect of the absorption
by NH in the wavelength range sampled by near-UV bandpasses such as the Johnson
$U$ (e.g. Marino et al. 2008), the Str\"omgren $u$ or even the Sloan $u$
bandpass (see Lardo et al. 2011), as also discussed by Sbordone et al. (2011)
and Carretta et al. (2011). We show in Fig.~\ref{f:ubuPIE} the results for
NGC~1851 using stars observed both by us and Grundahl: in {\it each} metallicity
group, the three components are nicely distributed along the RGB as if each
group were an individual GC.

The fraction of stars in the primordial, intermediate and extreme components
for NGC~1851 and the two MR and MP metallicity components are listed in
Table~\ref{t:quantenao}. Even considering the rather large errors from the Poisson
statistics, the fraction of primordial stars in the MR component appears
to be smaller than the P fraction in the MP component. These fractions
are computed using stars with both O and Na abundances (Carretta et al. 2009a).
If we use only Na abundances, regardless of O abundance (this by definition 
allows us only to separate first and second generation stars, since it is a 
criterion ``blind" to the separation between I and E stars) then the fractions
of primordial stars in the MR and MP components are still rather different:
$21\pm 5\%$ and $40\pm 10\%$, respectively.

\setcounter{table}{10}
\begin{table*}
\centering
\caption{Number of stars with both Na and O abundances and fraction of the 
Primordial, Intermediate, and Extreme components}
\begin{tabular}{lrrrr}
\hline
CASE       & N.stars&fraction &fraction &fraction \\
           & (O,Na) &P        &I	&E	  \\
           &GIR.+UVES	&component&component&component \\
\hline
NGC 1851 all& 95&  $31\pm ~6$ &  $63\pm ~8$ &  $ 6\pm 3$  \\
NGC 1851 MR & 61&  $20\pm ~6$ &  $72\pm ~11$ &  $ 8\pm 4$  \\
NGC 1851 MP & 34&  $44\pm ~11$ &  $53\pm ~12$ &  $ 3\pm 3$  \\
\hline
\end{tabular}
\label{t:quantenao}
\end{table*}

\begin{figure}
\centering
\includegraphics[bb=50 150 330 680,clip,scale=0.42]{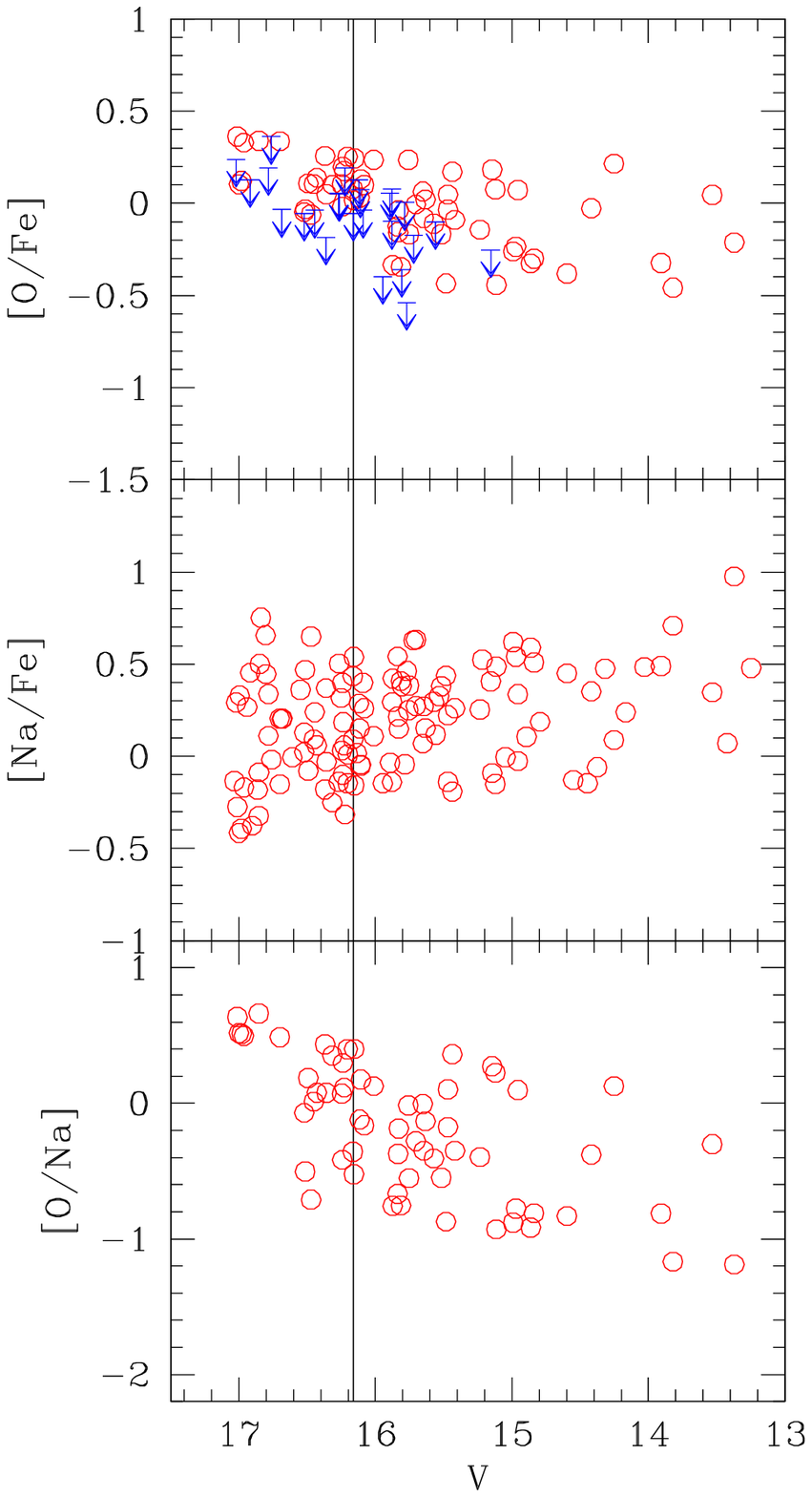}
\caption{[O/Fe], [Na/Fe] and [O/Na] ratios (upper, middle, and lower panel,
respectively) in RGB stars of NGC~1851 as a function of $V$ magnitude. The
RGB-bump luminosity level is indicated by a solid line. Upper limits in O are
indicated by arrows.}
\label{f:vona18}
\end{figure}

Fig.~\ref{f:vona18} shows the location of a slight change of the mean value of O
and Na abundances, just at the level of the bump on the RGB. We remind the
reader that our interpretation  (the change is caused by the mix of first and
second generation stars, each generation with its own slightly different He
content, see Salaris et al. 2006 and Bragaglia et al. 2010) is based on the
observation that  the ratios [Na/Fe] and [O/Fe] run essentially flat as a
function of luminosity in field stars (Gratton et al. 2000). While lighter
species (such as Li, C and N) can be mixed up, the ON and NeNa cycles require
higher temperatures: they are cycled in the inner layers of the H-burning shell
that cannot be reached by the extra-mixing processes even after the RGB-bump. In
the case of NGC~1851, where probably we could see the result of  two distinct
clusters, each with its own Na-O anticorrelation and P,I, and E components, 
we expect a  further smearing of the bump in the RGB luminosity function.

\begin{figure}
\centering
\includegraphics[scale=0.42]{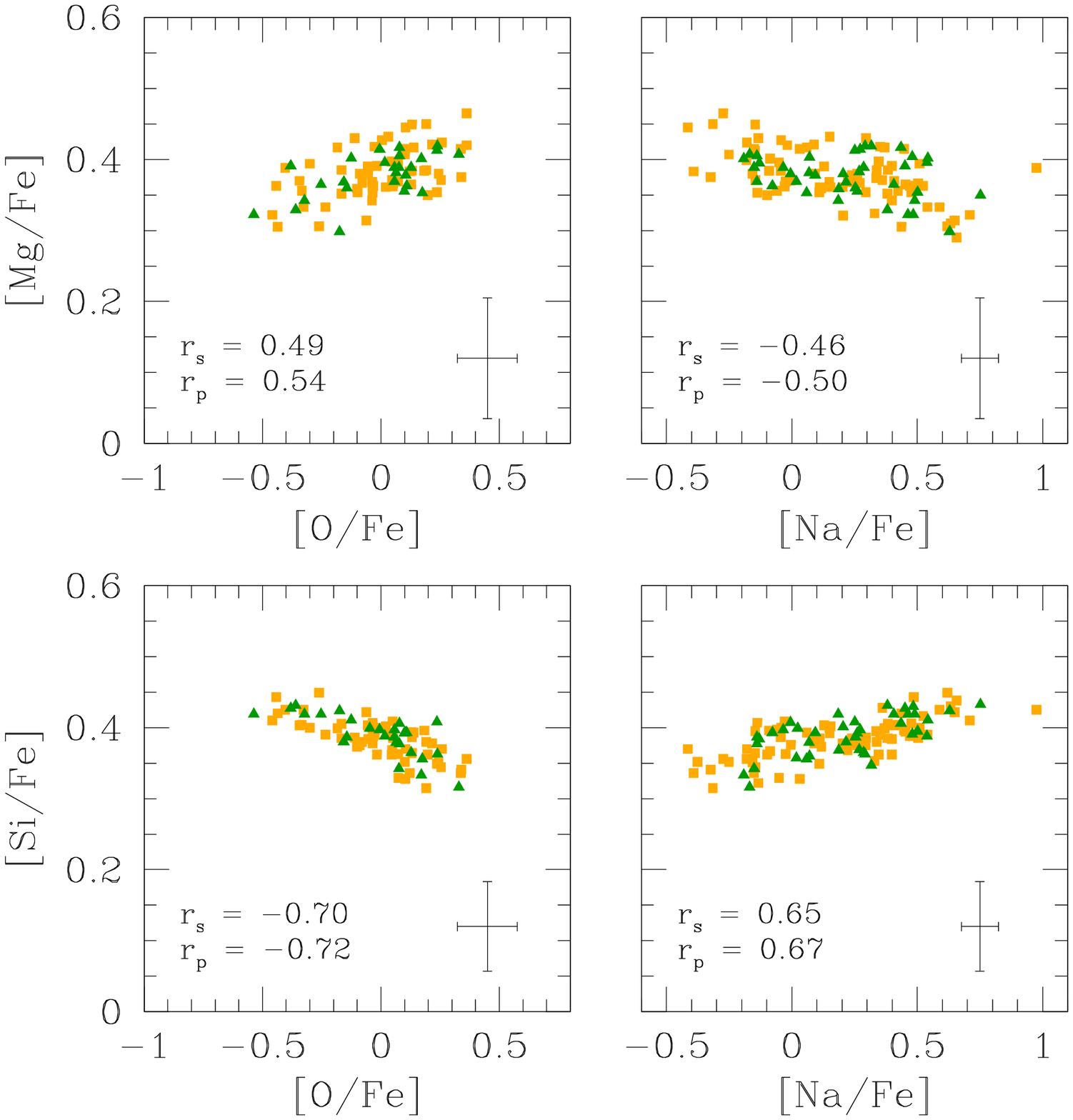}
\caption{Top panels: [Mg/Fe] ratios as a function of [O/Fe] and [Na/Fe] for
stars in NGC~1851 with GIRAFFE spectra. Bottom panels: [Si/Fe] ratios as a
function of [O/Fe] and [Na/Fe]. In each panel the internal error
bars and the Spearman rank and the Pearson linear regression coefficients are
indicated. Stars
are color coded according to their metallicity: green triangles for MP 
stars and orange squares for MR stars.}
\label{f:mgsinao18}
\end{figure}

Finally, we checked the other elements participating to the proton-capture
reactions, namely Mg and Si, using the large sample of stars with homogeneous
abundances from  GIRAFFE spectra. In NGC~1851 (Fig.~\ref{f:mgsinao18}) 
the abundances of Mg are  depleted as O is depleted and Na is enhanced, while
the opposite happens for Si.
The last occurrence is probably due to the leakage from the Mg-Al cycle on Si,
first discovered by Yong et al. (2005) in NGC~6752 and then found by Carretta et
al. (2009b) in stars of several clusters, although only in the small samples of 
RGB stars with UVES spectra. This is the first time that these relations
involving Si are observed in a very large sample of stars in an individual cluster.
The large statistics (85 stars with Mg, Si and O; about 120 stars with Mg, Si and
Na) makes all the (anti)correlations in Fig.~\ref{f:mgsinao18}
robust to a high level of confidence. Although the internal errors are not
negligible and the amount of variations in [Si/Fe] abundance ratios is not
dramatic (about 0.1-0.15 dex), these relations tell us that at least part
of the material from which the second generation stars formed was processed at a
temperature exceeding $\sim 65$ million K. This is the threshold for which the 
reaction producing $^{28}$Si becomes effective in the Mg-Al cycle (Arnould et al. 1999).

Separating the sample in the MR and MP components, the relations of
Fig.~\ref{f:mgsinao18} still hold in each sub-group.

\subsection{Nucleosynthesis from massive stars: $\alpha-$elements}

The elements involved in $\alpha-$capture reaction are the main signature of 
nucleosynthesis in massive stars, ending their lives as core-collapse supernovae.
Of course, here we are considering the level of {\it primordial} enrichment of
the gas from which the first stellar generation formed in the cluster.
Subsequently, the abundance of some elements, like Mg, and Si may be modified 
by proton-capture reactions, as
discussed in the previous section. Instead, no change should be observed for
heavier elements, e.g. Ca, between multiple generations in a GC.
The issue of chemical enrichment from massive stars has recently
received new interest after the proposal by Lee et al. (2009b) that chemical
pollution by type II SNe might be discerned in GCs hosting multiple populations.

\begin{figure}
\centering
\includegraphics[scale=0.42]{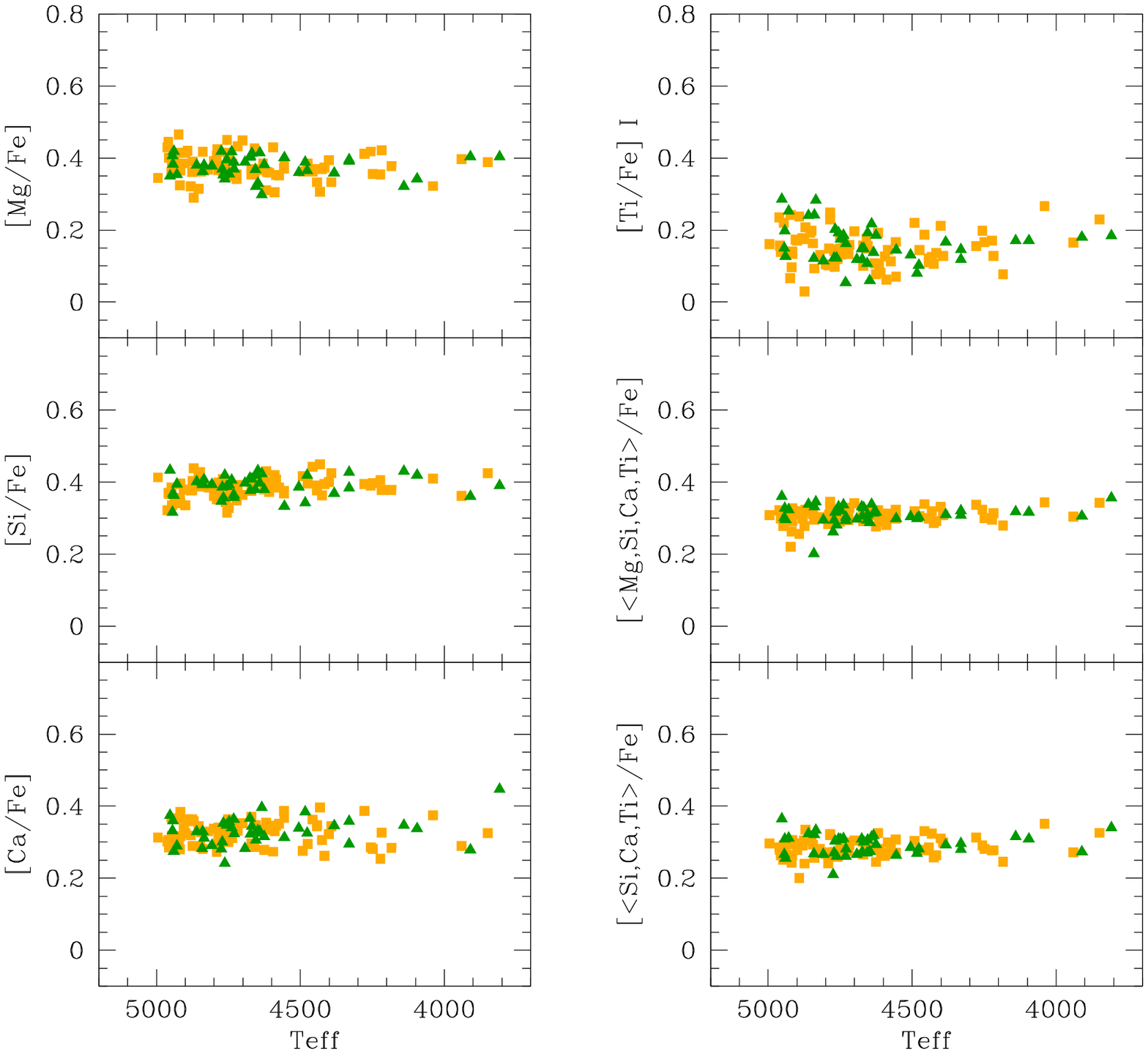}
\caption{Run of the abundance ratios of Mg, Si, Ca
(left panels), Ti {\sc i} (upper right panel) and two averages for the
$\alpha-$elements (last right panels) for stars in NGC~1851 with GIRAFFE
spectra. In each panel the color coding is as in Fig.~\ref{f:mgsinao18}.}
\label{f:alpm18}
\end{figure}

In Fig.~\ref{f:alpm18} we show the run of $\alpha-$elements in NGC~1851 as a
function of the temperature, using the large sample of stars with homogeneous
abundances from GIRAFFE spectra. The mean overabundance is quite normal
(Tab.~\ref{t:meanabu}) for cluster stars and the small observed scatter for Mg
and Si is compatible with the Mg-Al cycle producing only small changes to the
primordial level of Mg and Si, despite the noted anticorrelations and
correlations discussed in the previous Section. No distinct trend is observed as
a function of either effective temperature or metallicity, as evident by
the adopted color coding.

\begin{figure}
\centering
\includegraphics[scale=0.42]{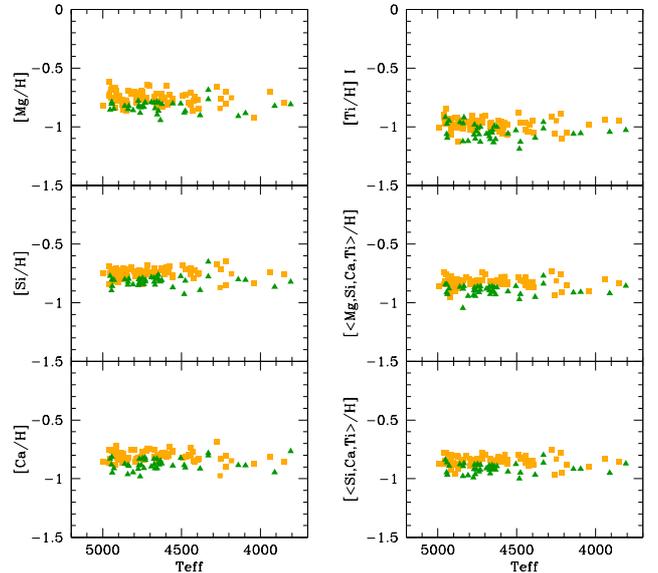}
\caption{As in Fig.~\ref{f:alpm18} but with element ratios all referred to H.
The color coding is as in the previous Figure.}
\label{f:alpm18H}
\end{figure}

However, the result is quite different if we plot the ratio with respect to the
H abundance, as in Fig.~\ref{f:alpm18H}: the absolute level of $\alpha-$elements
is larger in the MR component of NGC~1851 than in the MP component. The evidence
is somehow reduced for the case of Ti {\sc i} but is very clear for Mg, Si, Ca.

To put the comparison on a more quantitative ground, we list in
Table~\ref{t:mpmr} the average ratios measured for the MP and MR components
separately, using both the [el/Fe] and [el/H] abundance ratios. The value of
the variable $t$\ for the t-test is also listed in the Table and it is used to
assess whether the two averages are statistically different from each
other. We found that while the [el/Fe] mean values cannot be statistically
distinguished, the difference is significant with a high level of confidence
when we compare the average [el/H] ratios between the two components.

\setcounter{table}{11}
\begin{table}
\centering
\caption{Mean abundances from GIRAFFE spectra for stars in the MR and MP
component of NGC~1851}
\begin{tabular}{lccr}
\hline
                     &               &               & \\
Element              & MP component  & MR component  & t  \\
                     &n~~   avg~~  $rms$ &n~~	avg~~  $rms$ & \\        
\hline
$[$Mg/Fe$]${\sc i}   &39   +0.376 0.030 &77   +0.378 0.037 &  0.21\\
$[$Si/Fe$]${\sc i}   &40   +0.389 0.028 &77   +0.384 0.029 &  1.00\\
$[$Ca/Fe$]${\sc i}   &40   +0.329 0.038 &78   +0.326 0.033 &  0.48\\
$[$Ti/Fe$]${\sc i}   &39   +0.160 0.055 &77   +0.151 0.048 &  0.94\\
$[$Sc/Fe$]${\sc i}   &40   +0.025 0.059 &78   +0.029 0.057 &  0.38\\
$[$V/Fe$]${\sc i}    &38 $-$0.125 0.131 &75 $-$0.127 0.109 &  0.08\\
$[$Cr/Fe$]${\sc i}   &38   +0.023 0.110 &72   +0.064 0.114 &  1.83\\
$[$Co/Fe$]${\sc i}   &15 $-$0.027 0.073 &32 $-$0.020 0.071 &  0.28\\
$[$Ni/Fe$]${\sc i}   &40   +0.016 0.078 &77   +0.019 0.065 &  0.24\\
$[$Ba/Fe$]${\sc ii}  &32   +0.395 0.238 &69   +0.538 0.203 &  2.94\\
                     &                  &                  &      \\
$[$Mg/H$]${\sc i}    &39 $-$0.828 0.047 &77 $-$0.754 0.057 &  7.39\\
$[$Si/H$]${\sc i}    &40 $-$0.816 0.045 &77 $-$0.748 0.045 &  7.74\\
$[$Ca/H$]${\sc i}    &40 $-$0.876 0.049 &78 $-$0.807 0.047 &  7.41\\
$[$Ti/H$]${\sc i}    &39 $-$1.045 0.063 &77 $-$0.982 0.053 &  5.36\\
$[$Sc/H$]${\sc i}    &40 $-$1.181 0.078 &78 $-$1.104 0.076 &  5.14\\
$[$V/H$]${\sc i}     &38 $-$1.330 0.132 &75 $-$1.256 0.107 &  2.96\\
$[$Cr/H$]${\sc i}    &38 $-$1.183 0.113 &72 $-$1.069 0.129 &  4.79\\
$[$Co/H$]${\sc i}    &15 $-$1.231 0.085 &32 $-$1.149 0.076 &  3.21\\
$[$Ni/H$]${\sc i}    &40 $-$1.190 0.094 &77 $-$1.113 0.075 &  4.46\\
$[$Ba/H$]${\sc ii}   &32 $-$0.809 0.241 &69 $-$0.599 0.206 &  4.25\\

\hline
\end{tabular}
\label{t:mpmr}
\end{table}

Thus, the metal-rich component in NGC~1851 also shows a higher level of
$\alpha-$elements typically produced in core-collapse SNe.

In summary, these results are clearly telling us that the observed differences
in the primordial chemical composition between the MP and MR components in
NGC~1851 are due to core-collapse SNe and not to type Ia SNe, since the
$\alpha-$elements track the Fe abundance and there is no significant difference
in the [$\alpha$/Fe] ratios between the two populations. 

\begin{figure}
\centering
\includegraphics[scale=0.42]{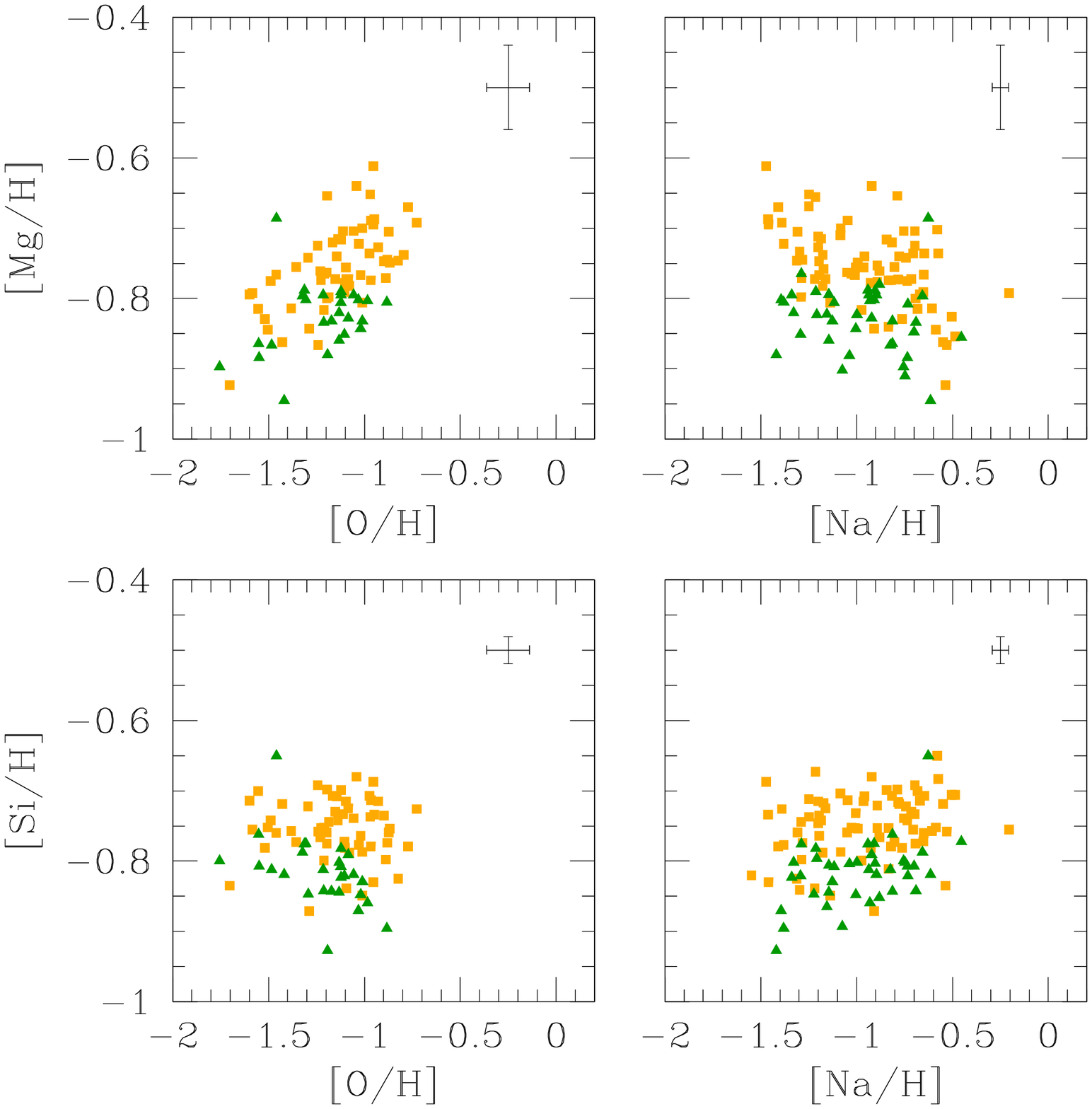}
\caption{Top panels: [Mg/H] ratios as a function of [O/H] and [Na/H] for
stars in NGC~1851 with GIRAFFE spectra. Bottom panels: [Si/H] ratios as a
function of [O/H] and [Na/H]. In each panel the internal error
bars are indicated. Stars are color coded according to their metallicity as in previous
figures.}
\label{f:mgsinaoH18}
\end{figure}

What happens to the (anti) correlations between Mg, Si and 
Na, O seen in the previous section? We plot again in
Fig.~\ref{f:mgsinaoH18} the stars of Fig.~\ref{f:mgsinao18}, color coded as
usual, but this time we use the [el/H] ratios. 

From this figure it is clear that the relations between elements involved in
proton-capture reactions still hold, {\it but they hold for each metallicity
component separately}: the only difference is that the level is shifted toward
higher values for the MR component. The existence of these neat 
(anti)correlation in each metallicity group again supports two 
originally distinct GCs: for what we know at present, this kind of trends among
light elements requires a precise chain of events and a nucleosynthesis only
possible in a GC formation scenario (see Carretta et al. 2010a).

Anyway, whatever the interpretation is, we confirm that a difference of [Ca/H]
does exist in NGC~1851. This difference was claimed by Lee et al. (2009b) from
the observed spread in their $hk$\ index.
A more detailed comparison between spectroscopic and photometric results will
be discussed below; however, as anticipated in Carretta et al. (2010c, their
Fig.4) using Str\"omgren photometry, stars along the RGB in NGC~1851 are not
segregated in metallicity [Fe/H], or [Ca/H], but
only according to their belonging to the first or second stellar generation.
There is little doubt that a spread of Ca is found in NGC~1851, but we found that
its mean level only correlates with [Fe/H] and not with [Na/O].

Our simple interpretation (the merger of two distinct clusters formed within a
large parent system) offers the advantage of explaining the observation in this
peculiar cluster, and at the same time to account for other GCs where a spread
of [Ca/H] values is excluded by high resolution spectroscopy (such as M~4,
NGC~6752 and other GCs with no metallicity spread, see Carretta et
al. 2010d).

\begin{figure}
\centering
\includegraphics[scale=0.42]{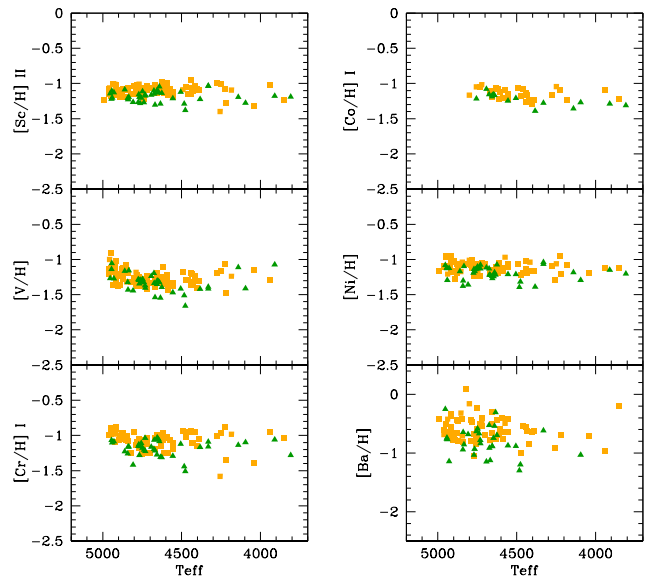}
\caption{Run of the abundance ratios of Fe-group elements Sc {\sc ii}, V, Cr, 
Co and Ni and of the neutron-capture element Ba {\sc ii} in stars of NGC~1851
with GIRAFFE spectra. All abundance ratios are plotted with respect to H.
The color coding used in previous figures is adopted.}
\label{f:fegm18H}
\end{figure}

The elements heavier than Ti, in particular the Fe-group elements (listed in
Tab.~\ref{t:fegroup18}), simply track the run of [Fe/H] in NGC~1851, hence
are slightly enhanced in the metal-rich component (Fig.~\ref{f:fegm18H}). 
Again, the difference between average values for the MR and MP components are
statistically significant when the [el/H] ratios are used (Table~\ref{t:mpmr}).

\subsection{Nucleosynthesis from less massive stars: neutron capture elements}

\subsubsection{$s-$ and $r-$process contributions}

Neutron capture elements can be produced by both slow $s-$process, and
by rapid $r-$process, where slow and rapid is for the $n-$capture with
respect to the $\beta-$decay. The $r-$process is active in (some?) core collapse
SNe, while the $s-$process is mainly active in the thermally pulsing phase
of intermediate-low mass stars (main component), although the light elements might also
be produced in massive stars (weak component). Hence, they have very
different timescales. Both $r-$ and $s-$processes 
likely contributed to make the $n-$capture elements observed in NGC~1851.
To estimate their relative importance, we compared the abundances of elements 
that in the Sun are mainly produced by the $s-$\ and $r-$process. In practice
(see also Carretta et al. 2010c), we considered the abundances of Ba and La 
as representative of $s-$process elements, and Eu to represent the 
contribution of the $r-$process. In the solar system the $r-$fraction of Eu is 
estimated to be more than 97\%, while it is $<40$\% for the other elements 
(for instance, it is $\sim 18$\% for Ba, see e.g. Simmerer et al. 2004). 

\begin{figure}
\centering
\includegraphics[scale=0.42]{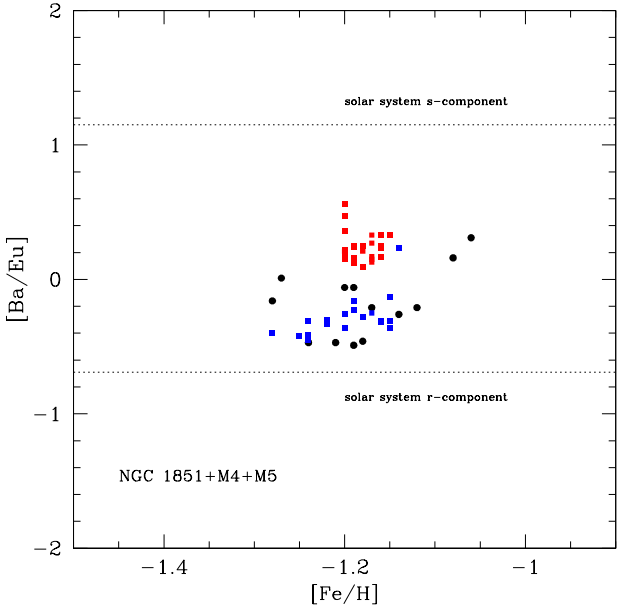}
\caption{Ratios [$s$/$r$] in stars of NGC~1851 (black filled circles), M~4 (red
squares, from Ivans et al. 1999) and M~5 (Ivans et al. 2001, blue squares).
Ba and Eu are assumed as representative of the level of $s$ and $r-$process
elements, respectively.}
\label{f:baeum4m5}
\end{figure}

The run of the [Ba/Eu] ratio with [Fe/H] in NGC~1851 is shown in Fig.~\ref{f:baeum4m5}. 
As a comparison, we also plotted the values available for two other 
GCs, well known to have a large difference in the abundance of neutron capture
elements (notably, those that in the Sun are mainly produced by $s-$process): 
M~4 (Ivans et al. 1999) and M~5 (Ivans et al. 2001). The range in [Ba/Eu] 
spanned by stars in NGC~1851 is large: about 0.80 dex, with 0.25 dex $r.m.s.$\ 
scatter. There are stars with [Ba/Eu] as high as that of M~4, whereas others 
have a ratio as low as that of M~5 ones.
 
In metal-poor stars of NGC~1851 the $s/r$\ ratio is lower than solar by about 
0.2 dex. It may be reproduced by assuming that about 1/3 of Ba is made by the 
$r-$process, and the rest by the $s-$process. An even larger fraction of other
elements that in the Sun are mainly due to the $s-$process (like e.g. Y and
Zr) is likely due to the $r-$process in the MP component of NGC~1851.
Over the metallicity range of NGC~1851, however, there seems to be a trend
for 
increasing Ba, La and Ce with respect to Eu as metallicity increases. For
the MR component, [Ba/Eu] is roughly solar. This is consistent with $\sim 4/5$
of the Ba of the MR component being due the $s-$process. In this case, the
$s-$process dominates also for other elements, including Y and Zr.

\subsubsection{Light and heavy s-elements}

Abundances of heavy elements might allow also to calculate the heavy 
(second peak) to light (first peak) $s$-process elements ratios for the 
stars in the sample. This ratio is important, since it can provide 
insight into the timescales of enrichment and the mechanisms responsible 
for it. 
In general, while it must be kept in mind that the neutron-density at the
site  of the $s-$process does in fact control it,  for a given neutron-density
(and metallicity) the heavy-to-light elements ratio does depend mostly on the
stellar mass,  with lower-mass stars ($\sim$1.5-2\,M$_{\odot}$) undergoing a
larger number of thermal pulses than $\sim$3\,M$_{\odot}$ stars (Travaglio et
al. 2004).   In fact, the main neutron source for the mass range under
consideration  (1.5-3\,M$_{\odot}$) is indeed the $^{13}$C($\alpha$,n)$^{16}$O,
whose sensitivity to temperature  is rather moderate, making the neutron-flux
very similar over the mass range. A high heavy-to-light ratio  should then be
related to a long timescale (even though overall metallicity  must also be taken
into consideration).

In order to obtain a meaningful evaluation of the heavy-to-light $s-$process
ratio, the $r$-process contribution must be subtracted from the abundances 
of the heavy elements. To this aim, we used the following approach.
We started by considering the solar $s-$ and $r-$ fractions
listed in Simmerer et al. (2004) for Y, Zr, Ba, La, and Eu.
We then assumed that the abundance ratios among different
elements with similar atomic weight produced by the $r-$ and $s-$process
are identical to the $r-$ and $s-$components measured in the Sun for all our 
sample stars; e.g., we have 
$\log{N_{\rm Eu_s}}/\log{N_{\rm Ba_s}}=$[$log{N_{\rm Eu_s}}/\log{N_{\rm Ba_s}}$]$_{\odot}$.
This is a reasonable assumption, since these elements are close to each other 
in the nucleosynthesis chain. On this basis, we derived the $s-$ and $r-$
contribution for Eu and Ba for all stars. We then assumed that the $r-$process contribution 
for all elements scales proportionally (that is, we assumed a universal
solar scaled $r-$process). This allows to determine the $r-$contribution for
all elements in all stars. We subtracted this contribution to
the abundances of all $n-$capture elements: the residual represents
our best guess for the $s$-process contribution for each element in each star.
This method provides reliable results insofar the $s-$contribution dominates
over the $r-$one. Otherwise, results are quite unstable, depending heavily
on observational errors.

\begin{figure}
\centering
\includegraphics[scale=0.42]{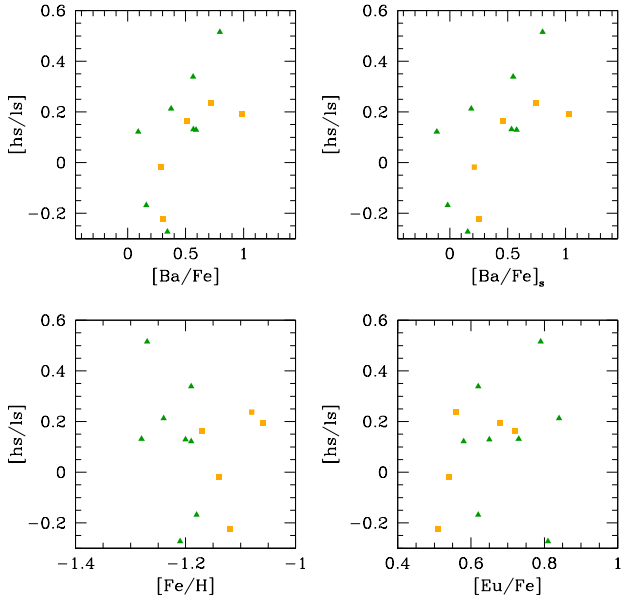}
\caption{[hs/ls] as function of [Fe/H], total Ba and Eu abundances and Ba
$s-$component abundance[Ba/Fe]$_s$. Closed and open symbols refer to the MR and
MP components, respectively}
\label{f:hsls}
\end{figure}

To reduce this concern, we took the average of Y$_s$ and Zr$_s$ to represent 
the light $s$-process elements and Ba$_s$ and La$_s$ for the heavy ones. 
Note that for a few stars the estimated $r-$component for Y and La accounted for
the entire measured  abundances for these elements (stars involved are \# 14080
and 43466 for both Y and La, and \# 39801 only for Y). In such cases we only
used Ba$_s$ for heavy-$s$ and Zr$_s$ for light-$s$.
The remaining elements were not used either because we conservatively deemed our
abundances not accurate enough (Ce) or because the large $r-$contribution makes
results unreliable. Plots for the behaviour of the ratio of heavy-to-light 
$s-$process elements as a function of [Fe/H], measured Ba and Eu abundances 
([Ba/Fe] and [Eu/Fe]) and [Ba/Fe]$_s$ are shown in Fig. \ref{f:hsls}.

Closed and open symbols are for the metal rich and the metal poor population 
respectively. The large scatter among the metal-poor population is expected,
since the $r-$contribution is typically large. In the metal-rich stars, for
which results are more reliable, the [hs/ls] ratios are typically quite high
\footnote{ We also find low and apparently constant Cu abundances in
NGC~1851. This is consistent with a negligible contribution by the weak
component of the $s-$process, since this element is thought to be
overproduced by this mechanism (Sneden et al. 1991).}.
This suggests a contribution by low-mass AGB 
stars (M$\sim$1.5-3\,M$_{\odot}$), at least for what concerns the 
high-metallicity population. However, it is unlikely that stars at the very lower 
end of this mass range (M$<2$M$_{\odot}$) have played a too large role in 
the chemical evolution of NGC~1851. In fact, these small mass stars do not deplete O; 
a large contribution by these stars would have diluted and even erased the Na-O 
anti-correlation. To maintain the Na-O anticorrelation, the contribution of the 
AGB stars that do not undergo HBB must be small with respect to those that do it.
As an exercise, we calculated the amount of mass ejected by small mass AGB (not 
undergoing HBB), and by more massive ones (undergoing HBB), assuming a Salpeter 
IMF. We found that adopting a threshold for HBB of M=3~M$_{\odot}$, then only stars 
with M$>$2.2\,M$_{\odot}$\ could contribute to the second generation of NGC~1851,
else the Na-O anticorrelation would be cancelled. Of course, this is a schematic
approach: there are likely stars which both experienced HBB and thermal pulses.
However, the qualitative argument that the contribution by low mass stars should
be small is still valid.

It is interesting to note that recently, D'Orazi et al. (2011) used a similar
approach to study the pattern of heavy elements in $\omega$ Cen. They found that
this requires AGB stars even more massive than those here considered as the site 
for the $s-$process observed in that cluster.

\subsubsection{Correlations between $n-$capture and $p-$capture element abundances}

\begin{figure*}
\centering
\includegraphics[scale=0.9]{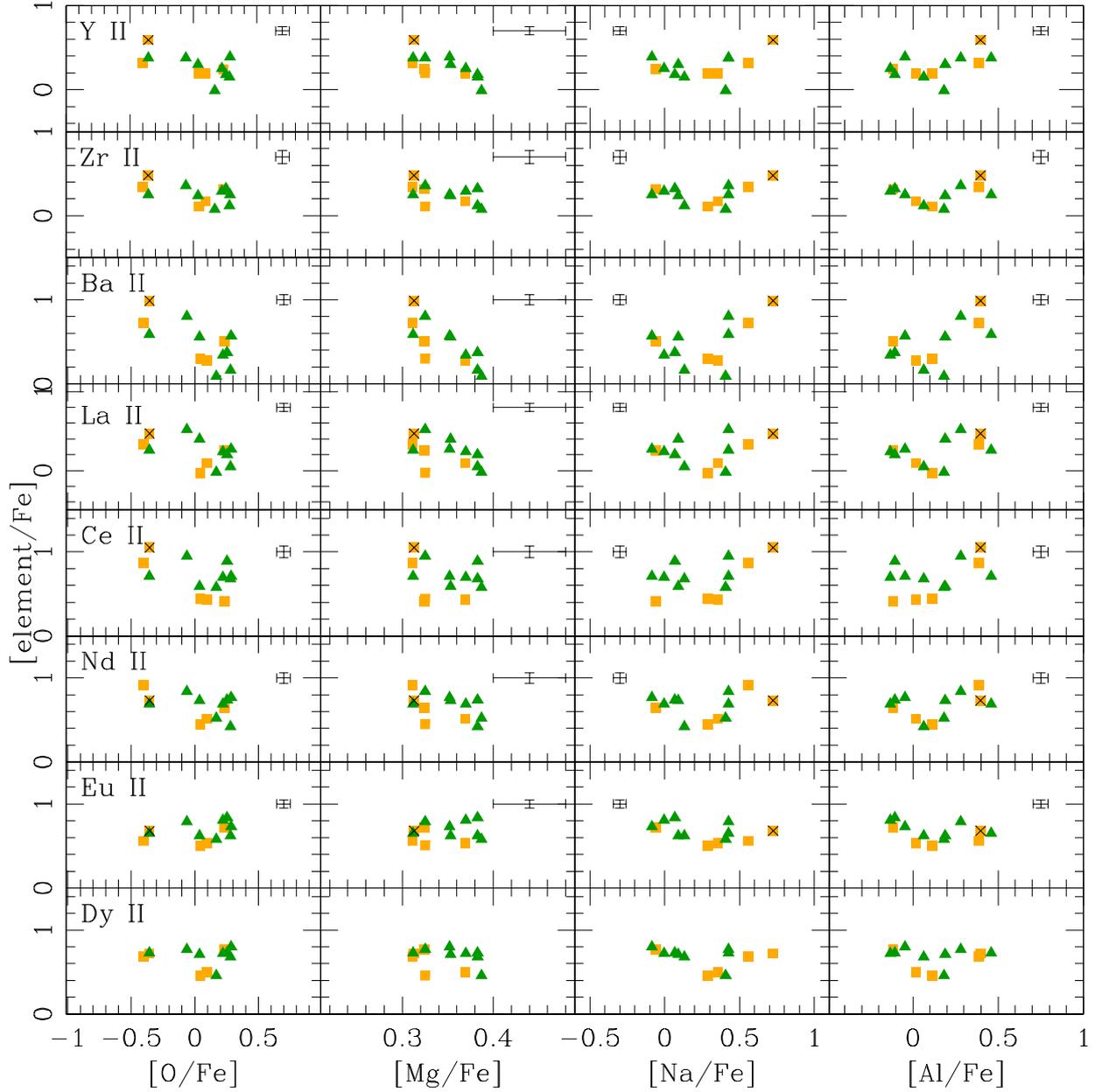}
\caption{Abundance ratios of neutron capture elements in stars with UVES
spectra, as a function of the proton capture elements O, Mg, Na, and Al. Stars
are separated in MR and MP components as in previous figures. The
star 32903 is indicated by a cross superimposed to the symbol. Error
bars indicate internal errors.}
\label{f:ncpcapture}
\end{figure*}

\begin{figure}
\centering
\includegraphics[bb=20 400 460 680,clip,scale=0.5]{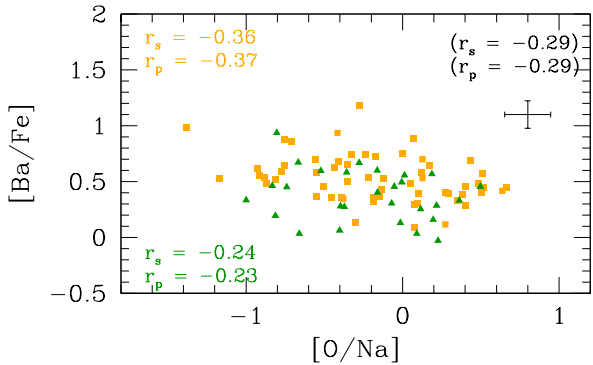}
\caption{[Ba/Fe] ratios as a function of [Na/O] for stars in NGC~1851 
with GIRAFFE spectra, separated according to their
sub-population. The Spearman rank coefficient $r_s$ and the Pearson linear
regression coefficient $r_p$ are indicated for the whole sample and
for both sub-populations.}
\label{f:banamg18a}
\end{figure}

\begin{table}
\centering

\small
\caption{Pearson's correlation coefficients, degree of freedom d.o.f., and level
of significance for relations of  neutron-capture vs proton-capture elements}
\begin{tabular}{llll}
\hline
element	           &   all            & MP		  &MR 		      \\
                   &  stars           &                   &                   \\
\hline     
Y~{\sc ii} vs O    &0.61 11 95-98\%   &  0.47 6  $<$90\% &  0.73 3 $<$90\% \\
Y~{\sc ii} vs Mg   &0.71 11    99\%   &  0.83 6     99\% &  0.56 3 $<$90\% \\
Y~{\sc ii} vs Na   &0.33 11 $<$90\%   &  0.11 6  $<$90\% &  0.72 3 $<$90\% \\
Y~{\sc ii} vs Al   &0.47 11 $<$90\%   &  0.25 6  $<$90\% &  0.73 3 $<$90\% \\

Zr~{\sc ii} vs O   &0.47 11 $<$90\%   &  0.24 6  $<$90\% &  0.61 3 $<$90\% \\
Zr~{\sc ii} vs Mg  &0.54 11 90-95\%   &  0.51 6  $<$90\% &  0.59 3 $<$90\% \\
Zr~{\sc ii} vs Na  &0.25 11 $<$90\%   &  0.14 6  $<$90\% &  0.46 3 $<$90\% \\
Zr~{\sc ii} vs Al  &0.24 11 $<$90\%   &  0.10 6  $<$90\% &  0.55 3 $<$90\% \\

Ba~{\sc ii} vs O   &0.69 11    99\%   &  0.56 6  $<$90\% &  0.78 3 $<$90\% \\
Ba~{\sc ii} vs Mg  &0.78 11    99\%   &  0.86 6     99\% &  0.70 3 $<$90\% \\
Ba~{\sc ii} vs Na  &0.44 11 $<$90\%   &  0.12 6  $<$90\% &  0.64 3 $<$90\% \\
Ba~{\sc ii} vs Al  &0.55 11    95\%   &  0.38 6  $<$90\% &  0.75 3 $<$90\% \\

La~{\sc ii} vs O   &0.48 11    90\%   &  0.43 6  $<$90\% &  0.66 3 $<$90\% \\
La~{\sc ii} vs Mg  &0.53 11 90-95\%   &  0.72 6  95-98\% &  0.56 3 $<$90\% \\
La~{\sc ii} vs Na  &0.21 11 $<$90\%   &  0.02 6  $<$90\% &  0.51 3 $<$90\% \\
La~{\sc ii} vs Al  &0.38 11 $<$90\%   &  0.24 6  $<$90\% &  0.59 3 $<$90\% \\

Ce~{\sc ii} vs O   &0.47 11 $<$90\%   &  0.09 6  $<$90\% &  0.94 3 98-99\% \\
Ce~{\sc ii} vs Mg  &0.23 11 $<$90\%   &  0.29 6  $<$90\% &  0.61 3 $<$90\% \\
Ce~{\sc ii} vs Na  &0.45 11 $<$90\%   &  0.12 6  $<$90\% &  0.86 3 90-95\% \\
Ce~{\sc ii} vs Al  &0.45 11 $<$90\%   &  0.07 6  $<$90\% &  0.93 3    98\% \\

Nd~{\sc ii} vs O   &0.45 11 $<$90\%   &  0.30 6  $<$90\% &  0.74 3 $<$90\% \\
Nd~{\sc ii} vs Mg  &0.46 11 $<$90\%   &  0.59 6  $<$90\% &  0.61 3 $<$90\% \\
Nd~{\sc ii} vs Na  &0.14 11 $<$90\%   &  0.09 6  $<$90\% &  0.44 3 $<$90\% \\
Nd~{\sc ii} vs Al  &0.29 11 $<$90\%   &  0.03 6  $<$90\% &  0.64 3 $<$90\% \\

Eu~{\sc ii} vs O   &0.29 11 $<$90\%   &  0.20 6  $<$90\% &  0.06 3 $<$90\% \\
Eu~{\sc ii} vs Mg  &0.14 11 $<$90\%   &  0.06 6  $<$90\% &  0.37 3 $<$90\% \\
Eu~{\sc ii} vs Na  &0.41 11 $<$90\%   &  0.35 6  $<$90\% &  0.21 3 $<$90\% \\
Eu~{\sc ii} vs Al  &0.38 11 $<$90\%   &  0.52 6  $<$90\% &  0.13 3 $<$90\% \\

Dy~{\sc ii} vs O   &0.06 11 $<$90\%   &  0.12 6  $<$90\% &  0.26 3 $<$90\% \\
Dy~{\sc ii} vs Mg  &0.28 11 $<$90\%   &  0.53 6  $<$90\% &  0.58 3 $<$90\% \\
Dy~{\sc ii} vs Na  &0.31 11 $<$90\%   &  0.42 6  $<$90\% &  0.02 3 $<$90\% \\
Dy~{\sc ii} vs Al  &0.38 11 $<$90\%   &  0.14 6  $<$90\% &  0.16 3 $<$90\% \\
\hline 
\label{t:corrtab}
\end{tabular}
\end{table}

In Fig.~\ref{f:ncpcapture} we investigated the
relation between neutron-capture and proton-capture elements in our sample,
using the full set of abundances available for the 13 stars with UVES spectra.
In Tab.~\ref{t:corrtab} we list the Pearson's correlation coefficient, the
number of the degree of freedom and the level of statistical confidence for the
linear fits made to the data in three cases: (i) using all the sample, (ii) and
(iii) separating the sample in MP and MR components.
Since in these plots a single star (32903) might drive part of the correlations,
we highlighted it in Fig.~\ref{f:ncpcapture} with a cross.

The first conclusion from these plots and the associated fits is that elements
like Eu and Dy, whose solar system abundances are almost totally due to the
$r-$process contribution, do not show any significant correlation with
proton-capture elements. Also the run of Nd with proton-capture elements is
essentially flat.

\begin{figure}
\centering
\includegraphics[scale=0.42]{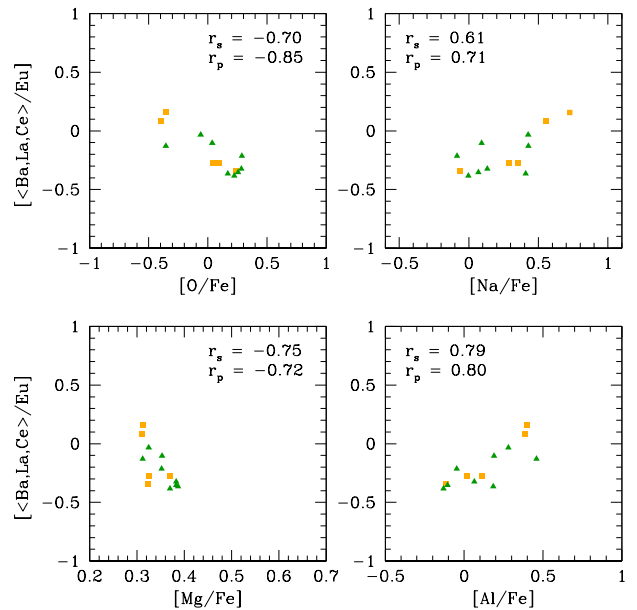}
\caption{The ratio [$<$Ba,La,Ce$>$/Eu] as a function of the proton capture elements
O. Na, Mg, Al for stars with UVES spectra of the MP (green symbols) and MR
(orange symbols) component in NGC~1851. Spearman rank correlation coefficient
and Pearson linear regression correlation coefficient are indicated in each
panel.}
\label{f:balaceeulightnoh}
\end{figure}

Yong and Grundahl (2008) noticed a possible correlation between the abundances
of the elements produced by $p-$capture an $n-$capture processes. A similar
result was also noticed by V10. On the whole, our much more extensive sample confirms
this finding; this is most clear when we use the large sample provided
by Giraffe spectra. While only a single Ba line (the one at 6141~\AA) could 
be measured on these spectra, the correlation between [Ba/Fe] and various indices 
related to the abundance of $p-$capture elements ([Na/Fe], the Str\"omgren $m_1$\ 
and $c_1$\ indices, the Lee $hk$\ index, etc.) is strong, at least for stars of the MR
component. The level of confidence is well above 99\% (see e.g., Fig.\ref{f:banamg18a}). While some trend with 
luminosity is also present, the correlations clearly hold also if we limit 
ourselves to the stars fainter than $V=15.5$, still with a very high level of
confidence. While all these results concern Ba, similar correlations are present 
also in the other mainly $s-$elements whose lines could be measured on the UVES spectra.
In Fig.~\ref{f:balaceeulightnoh}, to enhance our ``signal" we used an average
of Ba, La, Ce and we plotted the ratio [$<$Ba,La,Ce$>$/Eu] 
as a function of the abundances of $p-$capture elements for stars with high
resolution UVES spectra. In this Figure we report the 
Spearman rank and the Pearson's correlation coefficients relative to the whole 
sample of 13 stars; in this case, however, almost all the relations remain 
statistically significant (often at better than 99\% level of confidence) 
regardless that we separate the MP and MR components or even if the star 32903 
is dropped from the sample. These correlations are not due to individual
stars with anomalous values, as we see by plotting the
[Ba/Fe] ratio against the [Na/O] abundance for the much more numerous
stars with GIRAFFE spectra
Fig~\ref{f:banamg18a}. Again, the correlation is significant at a very high 
level of confidence for the MR stars, while the case for MP stars is more dubious.

Summarizing, we conclude that there is a strong evidence for a correlation
between $p-$capture and $n-$capture elements, at least for the MR component
of NGC~1851. Such a correlation is rarely found among GCs, and call for an
additional peculiarity of NGC~1851.

\subsubsection{Comparison with previous studies}

Other investigators have already published abundances of $n-$capture elements
in NGC~1851. Yong and Grundahl (2008) found large star-to-star variations 
in a sample of eight bright giants. In particular, they found that the abundances 
of Zr and La were anticorrelated with O and correlated with Al. While this
result agrees well with ours, we found that only those correlations 
involving La are actually statistically significant (here we assume that a level 
of confidence below 90\% for the Pearson's correlation coefficient implies 
that the associated correlation is not significant).

Five stars among our UVES sample (25859, 29719, 32903, 39801 and 41689) are in
common with V10, who report abundances of Fe, Na and O generally in reasonable
agreement with ours. However, our Y and Ba abundances are systematically larger
than V10, in some cases by as much as 0.3 and 0.6\,dex respectively.  Given that
V10 did neither list the adopted atmospheric parameters nor explicitly the atomic
parameter for these particular elements, we cannot attempt to trace the source
of this discrepancy. We only note that the abundances derived from the four Y and Ba
lines are in good agreement with each other in our data.

\section{Comparison of spectroscopy and photometry in NGC~1851}

NGC~1851 is clearly a complex globular cluster. Using our unprecedented large 
dataset of chemical abundances we may hope to have a better tagging of the 
observed broadening and/or splitting of photometric sequences and to have a
deeper insight on their origin.

\begin{figure*}
\centering
\includegraphics[scale=0.7]{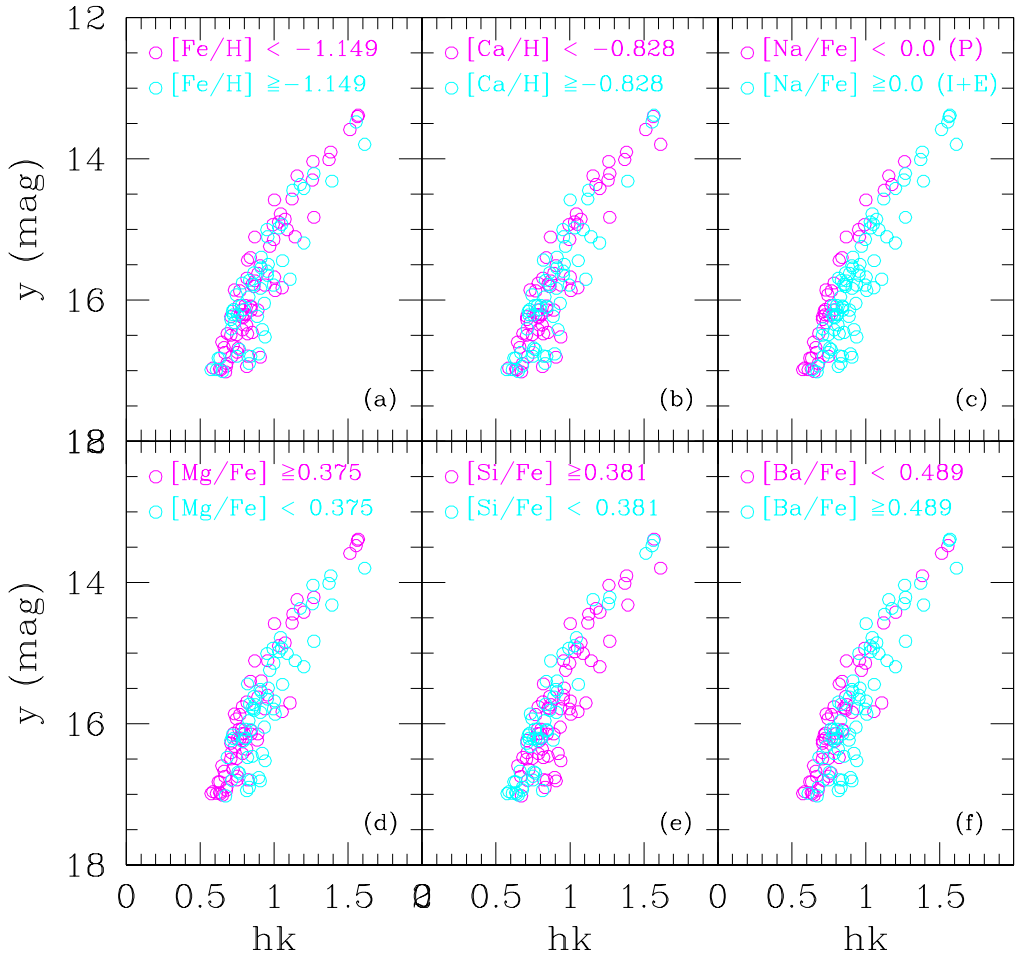}
\caption{RGB stars in common between our analysis and that by Lee et al.
(2009a) in the $y$-$hk$ plane from Str\"omgren photometry. In each panel stars
are color coded according to their abundances derived in the present study. The
elements and the separating values are indicated in each panel.}
\label{f:fighk}
\end{figure*}

In Fig.~\ref{f:fighk} we summarize in six CMDs the results of a first screening of
RGB stars in NGC~1851 using the $y$\ Str\"omgren magnitude and the $hk$\ index
kindly provided by prof. Jae-Woo Lee\footnote{We verified that the Lee
photometry agrees within about 0.05 mag with the Str\"omgren photometry by 
Grundahl, but in this case we adopt the first one since it is fully homogeneous
with the $hk$\ index ($hk=(Ca-b)-(b-y)$.}. In each panel we plot the giants of
NGC~1851 in common between the present study and the work of Lee and
collaborators (121 stars), colour-coded according to their abundances,
larger or lower than the average value for the common sample, listed in each
panel of Fig.~\ref{f:fighk}.

\begin{figure}
\centering
\includegraphics[bb=30 180 350 470,clip,scale=0.7]{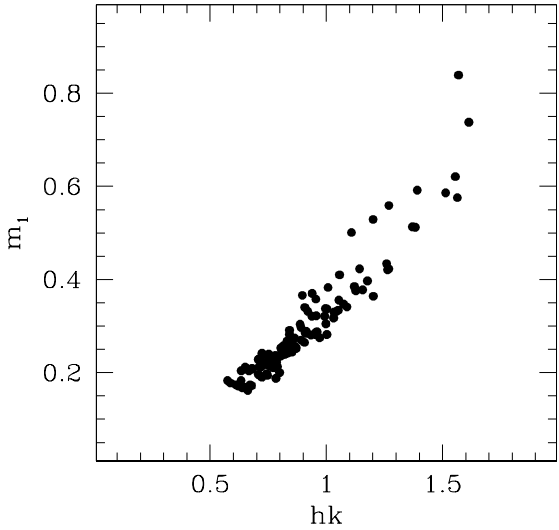}
\caption{Plot of $m_1$ and $hk$ for stars in NGC~1851 in common
between our work and Lee et al. (2009a).
Stars with higher values for $m_1$ and $hk$ are the brighter ones
of the sample.}
\label{f:figm1hk}
\end{figure}

Before discussing these plots, we notice that within NGC~1851, $hk$\ is extremely well correlated with the 
Str\"omgren index $m_1$ (see Fig.~\ref{f:figm1hk}). However, from our result, these indexes do 
neither appear to trace the metallicity of NGC~1851 stars (panel (a)), nor seem to 
be well correlated with a difference in [Ca/H], as shown in panel (b). 
Admittedly, as discussed in Section 4.2, the difference in [Fe/H] between the MR and 
MP components in NGC~1851 is small and might not be the best indicator to
separate stellar components in this peculiar cluster. However, the separation
between these components is real, as shown by their different radial 
distribution (Fig. 1 in Carretta et al. 2010c). Yet, the $hk$ index seems to be
blind to this difference.

\begin{figure}
\centering
\includegraphics[bb=30 180 350 470, clip,scale=0.7]{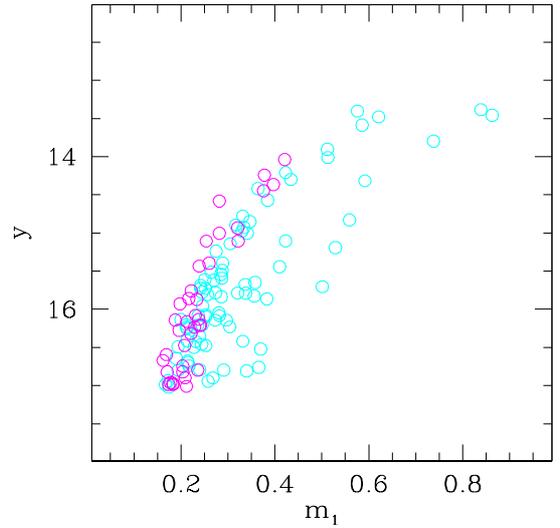}
\caption{Plot of $m_1$ vs $y$ for stars in NGC~1851, separated
in first (P, in magenta) and second (IE, in light blue) stars.}
\label{f:figm1y}
\end{figure}

On the other hand, in monometallic clusters $m_1$\ measures the strength of the near-UV CN bands, and it
is strongly correlated with N abundances in metal-rich clusters
(see Carretta et al. 2011). It may well be used to separate first and second
generation stars in NGC~1851 (see Fig.~\ref{f:figm1y}). So, it is not surprising
that also $hk$\ seems to be quite efficient in separating first and second
generation stars in NGC~1851, judging from panel (c) in Fig.~\ref{f:fighk}. In
this panel we did not adopt the average value for [Na/Fe] (+0.189 dex for stars
in common between us and Lee et al.), but we chose the value 0.0 that in
NGC~1851 divides first and second generation stars according to the definition
in Carretta et al. (2009a). The two stellar generations are separated very well
in the $y-hk$ plane. The same result would hold also using the ratio [O/Na],
although the separation is not so clean as in panel (c) due to the presence of
upper limits and/or larger errors in the O abundances.

The efficiency of $hk$\ to separate first and second generation stars is also 
supported by the CMDs in panels (d) and (e). Of course, separation of first and
second generation stars using Mg and Si is less accurate, because the spread in 
abundances is not very large. However, the striking evidence in these two 
panels is that the two groups are specularly selected, as expected due to the 
anticorrelation of Mg and Si.

The final panel (f) shows how stars are segregated along the split red giant
branch in NGC~1851 according to their Ba abundances. In this plane (but also in
the Str\"omgren $v,v-y$ CMD, where the split is even clearer) only Ba-rich stars
are found  on the reddest section of the RGB, whereas Ba-poor stars are mostly
(but not exclusively) restricted to the bluest side of the RGB. This finding
confirms and strengthens a similar result by V10, based on a much more limited sample.
At a given level of magnitude, all stars having high values of the $hk$\ index
are more Ba-strong than the average [Ba/Fe] value, although there is not a
one-to-one correspondence between Ba and $hk$.

\begin{figure}
\centering
\includegraphics[scale=0.42]{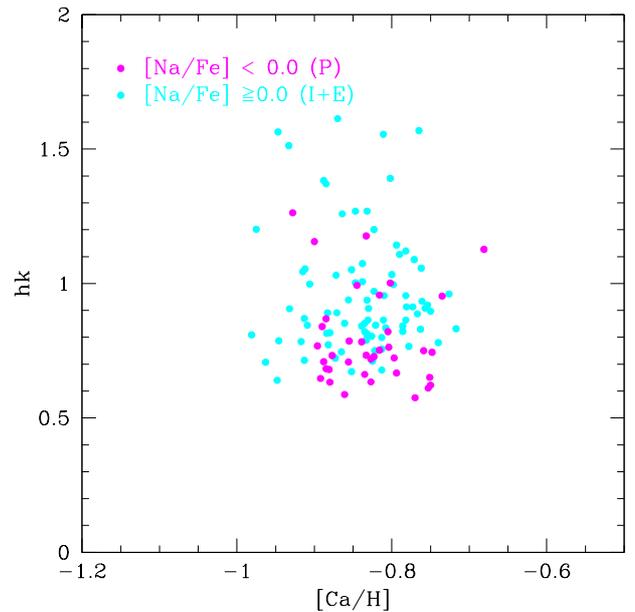}
\caption{$hk$ index by Lee and collaborators as a function of the [Ca/H] ratio
derived in the present study for giants in NGC~1851.}
\label{f:p8}
\end{figure}

Summarizing the results of this Section, we found that in NGC~1851 the $hk$\ index seems to
be related to the dichotomy first/second generation, but not to the
Ca abundance: in Fig.~\ref{f:p8} we also show the direct correspondence between
$hk$\ and our [Ca/H] values from spectroscopy, and we do not see any correlation.

In Carretta et al. (2009b) we demonstrated that in CMDs including the $u$ band,
second generation stars which are N-enhanced, as expected from the full
action of the complete CNO-cycle, are  spread out to the red, along the RGB. 
 The link between chemical composition
and observed colours in first and second generation stars is examined in
detail in another paper (Carretta et al. 2011), but from Fig.~\ref{f:ubuPIE}
and Fig.~\ref{f:fighk} we can conclude tentatively that $hk$\ separates stars
that are N-rich and N-poor, respectively. We notice that this result is
unexpected, given the definition of the Ca narrow band of Lee et al. (2009).

\begin{figure}
\centering
\includegraphics[scale=0.45]{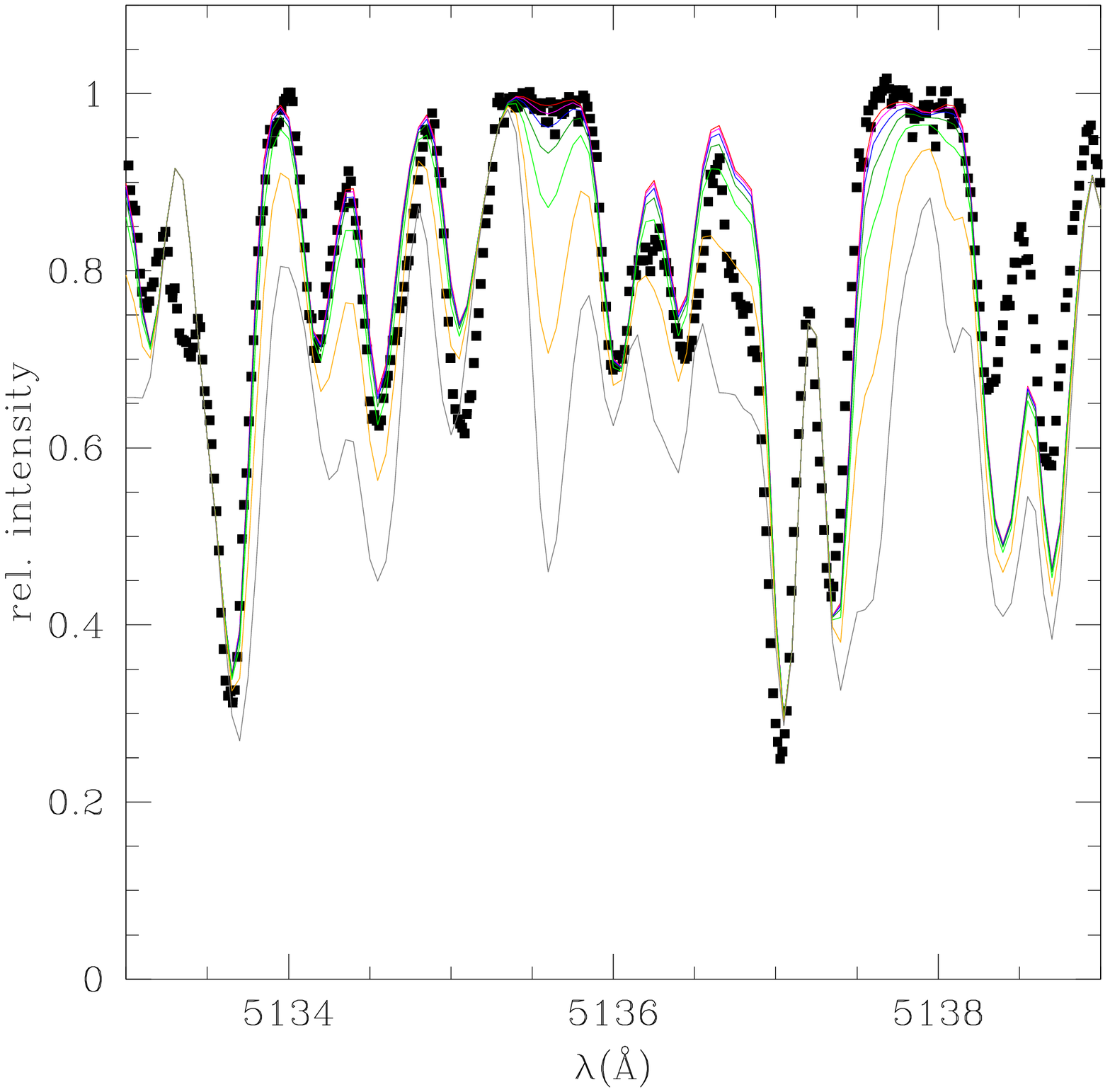}
\caption{Portion of the UVES spectrum of star 32903 in the C$_2$ region at
5135~\AA\ (filled squares) with synthetic spectra computed for various C
abundances ([C/Fe]=-0.5,-0.4,-0.3,-0.2,-0.1,0.0,+0.1).}
\label{f:c5630}
\end{figure}

\subsection{The effect of (neglecting) carbon}

Judging from panel (c) and (f) of Fig.~\ref{f:fighk}, there is a global
correspondence between Ba-rich and N-rich stars, but there are exceptions. This
is a significant difference with respect to the very tight sequences shown by V10 with
their limited sample, and it suggests that this is not the whole story.

\begin{figure*}
\centering
\includegraphics[bb=18 147 581 527, clip, scale=0.62]{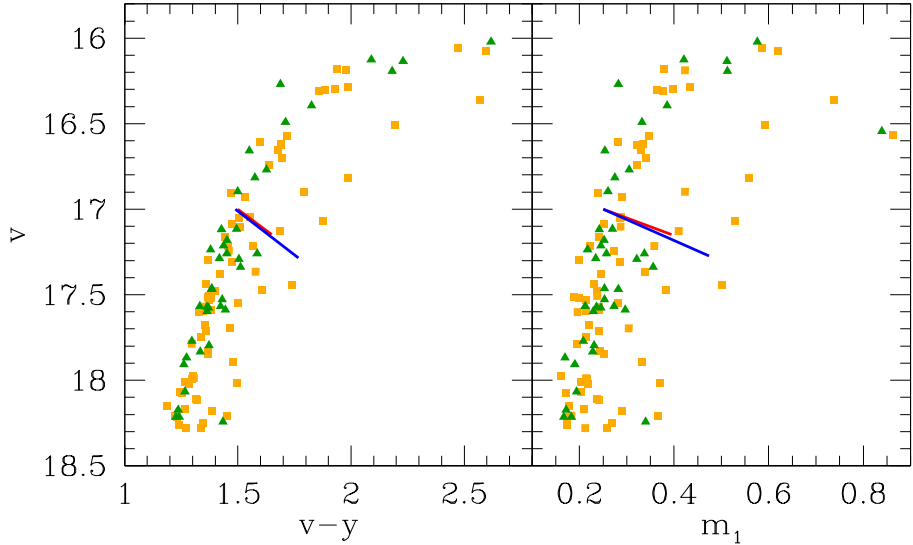}
\caption{Str\"omgren $v,v-y$ and $v,m_1$ for the MP and MR components in
NGC~1851. In both panels the red and blue lines indicate the effect
of a N and a C variation, respectively, on the stars' colours (see text).}
\label{f:cstrong}
\end{figure*}

Another possible player to be considered might be the carbon abundance. 
In another paper (Carretta et al. 2011) we studied in detail the effect
of variations of N and C abundances on the Str\"omgren photometry. Here we only
summarize in Fig.~\ref{f:cstrong} the main results relative to the C abundance
variations. 
Briefly, we used a model atmosphere with the parameters T$_{\rm eff}=4478$ K,
$\log g=1.4$ and [A/H]=-1.23 and we computed two synthetic spectra for a C-rich,
N-rich and a C-rich, N-poor case, respectively\footnote{In these
computations we adopted [O/Fe]=-0.1 dex for the C-rich, N-rich case and
[O/Fe]=+0.4 dex for the C-rich, N-poor case.}. Afterwards, we evaluated the
differences in magnitude with respect to the reference case of a C-normal,
N-poor star and we plot in the left and right panels of Fig.~\ref{f:cstrong} the
displacements due to variation in only N (red line) and also in C (blue line).

From this exercise it is easy to see that for N-poor stars, the impact of
varying C is modest. The main reason is that in these stars O $>$ C, and then
CN bands are not very strong. On the other hand, for N-rich stars there is a notable effect
when changing the C abundance, in particular in the $v$ filter. In turn, this
implies large variations in indices like $c_1$\ and $m_1$.

From these computations and from the CMDs in Fig.~\ref{f:cstrong} we could think that
stars on the redder sequence of NGC~1851 are quite good candidate to be C-rich
and N-rich (hence O-poor). On the other hand, the C-rich and N-poor stars (i.e.
O-rich) may not be separated from the other N-poor, C-normal stars in none
of the Str\"omgren indices. This might well explain the fact that the reddest RGB
sequence is only made by a trickling string of few stars: they might be not all the
C-rich stars present in the cluster, but only those that are also O-poor.
In general, it is likely that the C-rich stars in NGC~1851 would coincide with the Ba-rich stars, and
might correspond to our definition of MR population (Sect. 4.2), related to the
brighter SGB and to the red HB according to their number ratios.

However, the evidence supporting this view is currently limited to the
Str\"omgren photometry. We did not find evidence of C$_2$\ in our spectra,
even for the most Ba-rich (and also metal-rich) stars with UVES spectra 
(stars 32903 and 41689: see Fig.~\ref{f:c5630}). In the first case, we can even set a quite
stringent upper limit of [C/Fe]$<-0.4$, incompatible with the
assumption of C $\sim$ O which is basic in the scenario discussed in
this section. Of course, it is possible that C has been transformed into N
in these very bright giants, and that CN bands are very strong,
due to a very large excess of N. Unluckily, we could not prove this with
the present spectroscopic data.

\section{Discussion and conclusions}

There are several observations in NGC~1851 suggesting that two (originally) 
distinct objects may have concurred to form the currently observed  cluster: the
bimodal distribution of the HB, the double/split SGB, and the splitting recently
found on the RGB. To this, in Carretta et al. (2010c) we added a distinct radial
distribution, with the MP component more centrally concentrated.
When coupled with the existence of the Na-O anticorrelation in each
component on the RGB, although with different primordial/polluted ratios of
stars, this series of evidence points toward the existence of two merged clusters.
The results from the previous Section fit reasonably well a scenario where the two clusters
differ also for the average level of Ba (and probably also C) abundance, with
however additional complications that will be commented in the next subsection.

Our idea offers the advantage of accounting for a number of observations with
the minimum of assumptions. For most aspects, these essentially reduce to only one, 
that of an age difference of about 1~Gyr between the two putative clusters 
(since the metallicity difference - the other necessary assumption - is actually 
measured from our data). The age difference stems in part because we can exclude 
a significant contribution from He variations: the moderate extensions of both the Na-O and 
Al-Mg anticorrelations indicate that they must be not relevant. This minor role of He
variations is moreover confirmed by the constant luminosity level between the
red and the blue HB (see Cassisi et al. 2008).
An age difference of $\sim 1$~Gyr is also compatible with the higher level 
in $s-$process elements found for the MR component.

The combination of age and metallicity differences implies a difference of only
0.04~$M_\odot$ in the masses of stars currently on the RGB, which is compatible
with the HB but results into a negligible segregation of the two populations 
in the region of the cluster we observed. A primordial segregation between
the MP and MR populations might then have survived the following dynamical evolution.

Finally, there might also be a difference in the C content between the two
putative GCs; this is not only expected if the MR one is about 1 Gyr younger
(allowing the contribution of AGB stars with masses $<3~M_\odot$), but  could
also possibly account for the observed Str\"omgren colours. However, we did not
detect  any C$_2$ line in our spectra, which possibly contradicts the hypothesis
of a significant C enhancement (unless this is cancelled by evolutionary
effects).

In summary, most observations concerning NGC~1851 can be explained by the
existence of two sub-populations:
\paragraph{} A first one, including about 35-40\% of the stars, that behaves like a
small globular cluster, with a normal Na-O anticorrelations and normal features.
The only peculiar characteristic is maybe a number of primordial stars higher
than the average $\sim 30\%$ typical of GCs. This first population may be linked
to the faint SGB (Milone et al. 2008) and to the BHB and is more centrally
concentrated in NGC~1851, according to our RGB sample.

\paragraph{} A second population, formed by the remaining 60-65\% of stars, that
is represented by our MR component, more rich in Ba and $s-$process elements,
stronger, on average, in their $hk$ index, and with a behaviour in the Str\"omgren
photometric indices well explained by an overabundance of C. This population,
that is possible to identify with the bright SGB and the RHB, is not very
different in [Fe/H] or [Ca/H] from the first one, although it is clearly
more enriched in these elements. The most significant difference, however, seems to concern
the products from the third dredge-up, which implies an age difference large
enough to allow low mass stars to contribute. We also note that even if the content of C of
the MR population is perhaps higher than that in the MP population, the C+N+O sum
is probably not very different (0.1-0.2 dex, all concentrated on the C term),
because C $\ll$ (N+O) anyway. The contribution of low mass ($\sim 2-3
M_\odot$) stars is also required by the heavy-to-light ratio for the $s-$process
elements.

\subsection{The correlation between proton and neutron capture elements}

We confirm and extend the results found by Yong and Grundahl
(2008): in NGC~1851 the abundances of $s-$process dominated elements seems to be
strictly related to those of the light elements involved in $p-$capture
reactions.
The level of neutron capture elements increases for stars where the signature of
processing by H-burning at high temperature is more marked. While this signature
is very clean within the MR population, its presence is not obvious on the MP
one, depending on the membership of a couple of stars.

This observation bears a potential problem, depending on the mechanism 
assumed for the origin of the $s-$process elements (e.g. K\"appeler 1999). If,
as also suggested by the high heavy-to-light ratio,
they were forged by the {\it main} component (e.g. by low mass
stars with  $M_{ZAMS}\sim 1.5-3 M_\odot$ during the AGB phase), then the
evolutionary times of these stars are ranging from about 340 Myr up to about 2
Gyr (e.g. Schaller et al. 1992). 
However, if AGB stars with 4-8~$M_\odot$ are responsible
for the $p-$capture elements, the release of polluted matter from these
stars is expected as soon as about 40 Myr and as late as 160 Myr (the
timescale for the production of $p-$capture elements would be even shorter
were they produced by massive stars).
This timescale problem could be bypassed only if the gas cumulates
in a reservoir which does not form stars for several 
10$^8$~yrs\footnote{Note that this is not the long delay
of $\sim$~1 Gyr between MP and MR populations discussed at the beginning of
Section 7.}.

Such a time delay can perhaps be accommodated in models with a cooling flow as
tentatively proposed by D'Ercole et al. (2008) if e.g. a threshold in the mass
of gathering gas is necessary to trigger the formation of the second generation.
In addition, whether the correlation between $n-$ and $p-$capture elements
were continuous, dilution of the material of this reservoir with unprocessed
material would be required. Unluckily, our data are not accurate enough to favour or
dismiss this last hypothesis.

An alternative possibility would be that the same stars produced both $n-$ and $p-$capture
elements, as recently suggested by D'Antona et al. (2011) to explain a similar
problem for $\omega$~Cen. Such an hypothesis meets several difficulties, for
instance the lack of correlation between $n-$ and $p-$capture elements in most GCs.
Furthermore, the only known mechanism producing the $n-$capture elements in massive
stars (the weak component: Raiteri et al. 1991) is a secondary mechanism (then likely 
negligible in metal-poor stars) which preferentially produces the less massive among 
elements along the $s-$chain, and e.g. likely a significant fraction of Cu. At variance
with the recent case of M~22 (Marino et al. 2011), we do not find any sign of Cu
production within NGC~1851, and the heavy-to-light ratio is large. 

It then seems that the correlations between $p-$ and $n-$capture elements
in NGC~1851 observed by us and Yong and Grundahl (2008) put serious constraints 
to the scenario of pollution in GCs from AGB stars.

\subsection{Hunting for the ancestral galaxy}

The idea proposed by van den Bergh (1996), that GCs with a composite CMD are the
result of a merger of two parent clusters within the past equivalent of a
present-day dwarf spheroidal, offers several attractive hints to explain the
peculiarities observed in NGC~1851.

First, the N-body simulations by Makino et al. (1991) show that orbital angular
momentum is transferred at earlier stages to the escaping particles: this
process favours the shrinkage of the orbit, hence the merger. Nowadays, we are
aware that a large loss of stars is expected at early phases from a
proto-GC, since this is required by the observed composition of first and second
generation stars (see e.g. Carretta et al. 2010a; Conroy 2011). Should the early
cluster evolution occurs in the environment of a dSph, this large mass loss may
actually help the merger process.

If the clusters were in the same dwarf galaxy, they might have both spiralled to the
centre (by dynamical friction) and then merged in the currently observed object.
A necessary condition for this chain of events is that the dSph must have survived
at least for a few Gyr to the interaction with our Galaxy.
Cases for dSphs hosting a system of GCs of their own are well known (Fornax, and
the disrupting Sagittarius dwarf).

On the other hand, the different density distribution of MP and MR stars
suggests that the merging of the two clusters occurred not too long in the past,
less than a few Gyr ago. Since this merging should have happened before the
putative dwarf spheroidal dissolved, it is possible that the 
remains of this ancestral, parent galaxy can still be traced around
NGC~1851. Recently, a study by Olszewski et al. (2009) found that the cluster
is embedded in a halo 500 pc across, corresponding to about 0.1\% of the
dynamical mass of NGC~1851. The two populations - as represented by twin MSs and
SGBs - are traced up to more than 2.85 tidal radii from the cluster centre,
incidentally contradicting the finding by Zoccali et al. (2009) on a strong radial
gradient for the SGB. More recently, Carballo-Bello \& Martinez-Delgado (2010)
detected with the ESO-WFI a clearly traceable MS of extra-tidal stars around
NGC~1851 and NGC~1904. It is interesting to note that both studies do not find
evidence for collimated structure and both conclude against the evidence of
tidal tails: NGC~1851 is simply embedded in a smooth, giant cloud of tidal 
debris. 

According to Carballo-Bello \& Martinez-Delgado, it is tempting to associate
this overdensity to the debated Canis Major dwarf system: this conclusion would
be corroborated by the observation that the old and metal-poor population of the
debris cover a sky-projected distance of about 3.3 kpc, similar to the distance
between NGC~1851 and NGC~1904. Both clusters are considered (with NGC~2298 and
NGC~2808) to be part of the Canis Major system (Martin et al. 2004).
On the other hand, Olszewski et al. (2009) conclude that from existing dynamical
models it is not possible to discriminate between isolated cluster evaporation
and destruction of a dwarf galaxy, for the origin of the observed giant halo
around NGC~1851.

Is this cluster an exception? NGC~1851 shares several characteristics with another,
more metal-poor cluster: M~22. Marino et al. (2009, 2011) found two sub-populations
differing both in Fe and Ba, in this cluster, with the abundance of $s-$process
elements correlated with the iron abundance. The metallicity spread they found
(0.10 dex $r.m.s.$ scatter) is not much different from the values found by high
resolution spectra by us (0.07 dex) or by Yong and Grundahl (2008; 0.084 dex)
for NGC~1851, and the metal-rich population is also Ca-rich with respect to the
metal-poor one.

Finally, we also note that the merger hypothesis was recently considered by Lane et
al. (2010a) to account for some kinematic features not well explained for the 
globular cluster 47~Tuc. For example, Lane et al. (2010b) note that a merger
between two  clusters with similar metallicity ``would have increased the
total luminosity of 47~Tuc without altering its $M/L_V$\ significantly": this
results in the cluster being an outlier in their $M/L_V$ vs $M_V$ relation.
However, a two component model with populations having two distinct kinematics 
is favoured in the case of 47~Tuc, whereas no significant kinematic 
difference can be established between the two populations in NGC~1851, apart
from the observed radial segregation of stars on the RGB.

It is thus just possible that several channels are viable for the formation of
different ``flavours" of globular clusters. Beside the large, massive GCs
spiralling in the nucleus of their own galaxy, finding only a nuclear diffuse
component, as is the case of M~54 and perhaps $\omega$~Cen (e.g. Carretta et al.
2010b), possibly another channel is feasible: the merger of two GCs in the 
same parent galaxy before the disruption within a much larger galaxy.

However, more observations are needed to settle the case of NGC~1851.
In the future, we plan to analyse CN and CH low resolution spectra of SGB stars
in each of the two SGBs found in NGC~1851 to investigate the role of C and N
abundances. We also intend to test our
hypothesis of a cluster merger by exploring the Na-O anticorrelation in both
the BHB and RHB of this intriguing cluster.

\begin{acknowledgements}
We thank Jae-Woo Lee for sending us his data on NGC~1851 in advance of
publication, Antonio Sollima for useful discussions on tidal remains around the
cluster, Santino Cassisi for estimates of the maximum temperature along the HB
for M~54 and $\omega$ Cen and Yazan Momany for sending us his unpublished
photometry of NGC~1851. We also thank the referee, whose suggestions helped to
improve the paper. This publication makes use of data products from the Two
Micron All Sky Survey, which is a joint project of the University of
Massachusetts and the Infrared Processing and Analysis Center/California
Institute of Technology, funded by the National Aeronautics and Space
Administration and the National Science Foundation.  This research has been
funded by PRIN MIUR 20075TP5K9, and by PRIN INAF "Formation and Early Evolution
of Massive Star Clusters". This research has made use of the SIMBAD database,
operated at CDS, Strasbourg, France and of NASA's Astrophysical Data System.
\end{acknowledgements}

\begin{appendix}

\section{Error estimates}

We refer the reader to the analogous Appendices in Carretta et al. (2009a,b) for
a detailed discussion of the procedure adopted for error estimates. 
In the following we only provide the main tables of sensitivities of abundance
ratios to the adopted errors in the atmospheric parameters and $EW$s and the
final estimates of internal and systematic errors for all species analysed from
UVES and GIRAFFE spectra of stars in NGC~1851.

The sensitivities of derived  abundances on the adopted atmospheric parameters 
were obtained by repeating our abundance analysis by changing only one 
atmospheric parameter each time for $all$\ stars in NGC~1851 and taking the 
average value of the slope change vs. abundance. This exercise was done separately 
for both UVES and GIRAFFE spectra.

We notice that when estimating the contribution to internal errors due to EWs
and $v_t$, the values usually adopted (determined from the scatter of abundances 
from individual lines) are overestimated, because regularities in
the data are not taken into account. These regularities are due to
uncertainties in the $gf-$values, unrecognised blends with adjacent lines,
not appropriate positioning of the continuum, etc. They show up in uniform deviations
of individual lines from average abundances for each star. By averaging
over all stars the residuals of abundances derived from individual lines with 
respect to the average value for each star, we estimated that some 36\% of 
the total variance in the Fe abundances from individual lines is due to 
systematic offsets between different lines, which repeat from star-to-star. 
For about 30\% of the line, these offsets have trends with temperature
significant at about 2~$\sigma$ level. However, we found that the additional 
fraction of variance that can be explained by these trends is very small, and 
we can neglect it. We conclude that when considering star-to-star variations 
(internal errors, according to our denomination), the errors in EWs and $v_t$ 
should be multiplied by 0.8. 

The amount of the variations in the atmospheric parameters is shown in the
first line of the headers in Table~\ref{t:sensitivityu18},
and Table~\ref{t:sensitivitym18}, whereas the resulting 
response in abundance changes of all elements (the sensitivities) are shown 
in columns from 3 to 6 of these tables.

\begin{table*}
\centering
\caption[]{Sensitivities of abundance ratios to variations in the atmospheric
parameters and to errors in the equivalent widths, and errors in abundances for
stars in NGC 1851 observed with UVES}
\begin{tabular}{lrrrrrrrr}
\hline
Element     & Average  & T$_{\rm eff}$ & $\log g$ & [A/H]   & $v_t$    & EWs     & Total   & Total      \\
            & n. lines &      (K)      &  (dex)   & (dex)   &kms$^{-1}$& (dex)   &Internal & Systematic \\
\hline        
Variation&             &  50           &   0.20   &  0.10   &  0.10    &         &         &            \\
Internal &             &   4           &   0.04   &  0.06   &  0.04    & 0.070   &         &            \\
Systematic&            &  36           &   0.06   &  0.03   &  0.01    &         &         &            \\
\hline
$[$Fe/H$]${\sc  i}& 82 &   +0.039      &   +0.024 &  +0.006 & $-$0.034 & +0.008  &0.017    &0.033      \\
$[$Fe/H$]${\sc ii}&  9 & $-$0.063      &   +0.107 &  +0.036 & $-$0.015 & +0.023  &0.039    &0.064      \\
$[$O/Fe$]${\sc  i}&  2 & $-$0.019      &   +0.060 &  +0.028 &   +0.034 & +0.048  &0.054    &0.071      \\
$[$Na/Fe$]${\sc i}&  4 &   +0.017      & $-$0.056 &$-$0.039 &   +0.016 & +0.034  &0.043    &0.074      \\
$[$Mg/Fe$]${\sc i}&  3 & $-$0.007      & $-$0.017 &$-$0.008 &   +0.019 & +0.039  &0.040    &0.011      \\
$[$Al/Fe$]${\sc i}&  2 &   +0.012      & $-$0.029 &$-$0.014 &   +0.031 & +0.048  &0.051    &0.066      \\
$[$Si/Fe$]${\sc i}&  9 & $-$0.055      &   +0.021 &  +0.008 &   +0.025 & +0.023  &0.026    &0.043      \\
$[$Ca/Fe$]${\sc i}& 17 &   +0.037      & $-$0.037 &$-$0.020 & $-$0.022 & +0.016  &0.023    &0.030      \\
$[$Sc/Fe$]${\sc ii}& 8 &   +0.053      & $-$0.028 &$-$0.006 & $-$0.019 & +0.024  &0.026    &0.045      \\
$[$Ti/Fe$]${\sc i}&  9 &   +0.058      & $-$0.030 &  +0.008 & $-$0.003 & +0.023  &0.024    &0.047      \\
$[$Ti/Fe$]${\sc ii}& 1 &   +0.042      & $-$0.021 &$-$0.003 &   +0.009 & +0.068  &0.068    &0.034      \\
$[$V/Fe$]${\sc i} &  8 &   +0.072      & $-$0.018 &$-$0.003 &   +0.007 & +0.024  &0.025    &0.057      \\
$[$Cr/Fe$]${\sc i}&  3 &   +0.029      & $-$0.025 &$-$0.017 &   +0.018 & +0.039  &0.041    &0.026      \\
$[$Cr/Fe$]${\sc ii}&12 &   +0.019      & $-$0.030 &$-$0.018 &   +0.003 & +0.020  &0.023    &0.031      \\
$[$Mn/Fe$]${\sc i}&  3 &   +0.021      & $-$0.015 &$-$0.009 & $-$0.001 & +0.039  &0.040    &0.025      \\
$[$Co/Fe$]${\sc i}&  5 & $-$0.009      &   +0.002 &  +0.002 &   +0.024 & +0.030  &0.032    &0.017      \\
$[$Ni/Fe$]${\sc i}& 29 & $-$0.018      &   +0.017 &  +0.007 &   +0.015 & +0.013  &0.015    &0.017      \\
$[$Cu/Fe$]${\sc i}&  1 &   +0.011      & $-$0.074 &  +0.044 & $-$0.084 & +0.068  &0.081    &0.032      \\
$[$Y/Fe$]${\sc ii}&  5 &   +0.041      & $-$0.007 &  +0.008 & $-$0.085 & +0.030  &0.046    &0.052      \\
$[$Zr/Fe$]${\sc i} & 4 &   +0.087      & $-$0.020 &$-$0.018 &   +0.027 & +0.034  &0.038    &0.074      \\
$[$Zr/Fe$]${\sc ii}& 1 &   +0.092      & $-$0.099 &$-$0.036 & $-$0.087 & +0.068  &0.082    &0.079      \\
$[$Ba/Fe$]${\sc ii}& 4 &   +0.083      & $-$0.050 &  +0.004 & $-$0.120 & +0.034  &0.060    &0.096      \\
$[$La/Fe$]${\sc ii}& 2 &   +0.110      & $-$0.180 &$-$0.036 & $-$0.035 & +0.048  &0.066    &0.106      \\
$[$Ce/Fe$]${\sc ii}& 1 &   +0.063      & $-$0.027 &$-$0.007 & $-$0.005 & +0.068  &0.069    &0.072      \\
$[$Nd/Fe$]${\sc ii}& 2 &   +0.073      & $-$0.043 &$-$0.011 & $-$0.105 & +0.048  &0.065    &0.069      \\
$[$Eu/Fe$]${\sc ii}& 2 &   +0.110      & $-$0.087 &  +0.014 &   +0.010 & +0.048  &0.053    &0.089      \\
$[$Dy/Fe$]${\sc ii}& 1 &   +0.078      & $-$0.026 &$-$0.002 & $-$0.005 & +0.068  &0.069    &0.066      \\
\hline
\end{tabular}
\label{t:sensitivityu18}
\end{table*}

\begin{table*}
\centering
\caption[]{Sensitivities of abundance ratios to variations in the atmospheric
parameters and to errors in the equivalent widths, and errors in abundances for
stars in NGC 1851 observed with GIRAFFE}
\begin{tabular}{lrrrrrrrr}
\hline
Element     & Average  & T$_{\rm eff}$ & $\log g$ & [A/H]   & $v_t$    & EWs     & Total   & Total      \\
            & n. lines &      (K)      &  (dex)   & (dex)   &kms$^{-1}$& (dex)   &Internal & Systematic \\
\hline        
Variation&             &  50           &   0.20   &  0.10   &  0.10    &         &         &            \\
Internal &             &   4           &   0.04   &  0.03   &  0.14    & 0.111   &         &            \\
Systematic&            &  51           &   0.06   &  0.06   &  0.02    &         &         &            \\
\hline
$[$Fe/H$]${\sc  i}& 31 &   +0.056      & $-$0.006 &$-$0.002 & $-$0.035 & +0.020  &0.043    &0.057      \\
$[$Fe/H$]${\sc ii}&  3 & $-$0.032      &   +0.091 &  +0.024 & $-$0.012 & +0.064  &0.057    &0.043      \\
$[$O/Fe$]${\sc  i}&  1 & $-$0.045      &   +0.090 &  +0.033 &   +0.036 & +0.111  &0.100    &0.059      \\
$[$Na/Fe$]${\sc i}&  3 & $-$0.013      & $-$0.029 &$-$0.013 &   +0.023 & +0.064  &0.058    &0.035      \\
$[$Mg/Fe$]${\sc i}&  2 & $-$0.022      & $-$0.003 &$-$0.004 &   +0.021 & +0.078  &0.067    &0.023      \\
$[$Si/Fe$]${\sc i}&  7 & $-$0.053      &   +0.034 &  +0.009 &   +0.029 & +0.042  &0.048    &0.055      \\
$[$Ca/Fe$]${\sc i}&  5 &   +0.003      & $-$0.030 &$-$0.005 & $-$0.002 & +0.050  &0.040    &0.010      \\
$[$Sc/Fe$]${\sc ii}& 5 & $-$0.058      &   +0.087 &  +0.029 &   +0.013 & +0.050  &0.047    &0.065      \\
$[$Ti/Fe$]${\sc i}&  4 &   +0.020      & $-$0.004 &$-$0.010 &   +0.015 & +0.056  &0.048    &0.021      \\
$[$V/Fe$]${\sc i} &  6 &   +0.030      & $-$0.001 &$-$0.010 &   +0.020 & +0.045  &0.043    &0.033      \\
$[$Cr/Fe$]${\sc i}&  4 &   +0.003      & $-$0.002 &$-$0.008 &   +0.026 & +0.056  &0.053    &0.012      \\
$[$Co/Fe$]${\sc i}&  1 &   +0.004      &   +0.019 &  +0.005 &   +0.025 & +0.111  &0.093    &0.014      \\
$[$Ni/Fe$]${\sc i}&  7 & $-$0.015      &   +0.024 &  +0.005 &   +0.018 & +0.042  &0.040    &0.018      \\
$[$Zr/Fe$]${\sc i}&  2 &   +0.066      & $-$0.013 &$-$0.014 &   +0.031 & +0.078  &0.072    &0.086      \\
$[$Ba/Fe$]${\sc ii}& 1 & $-$0.043      &   +0.041 &  +0.044 & $-$0.030 & +0.111  &0.097    &0.051      \\
\hline
\end{tabular}
\label{t:sensitivitym18}
\end{table*}

\end{appendix}

\end{document}